\documentclass[twocolumn,trackchanges,floatfix]{aastex631}
\usepackage{amsmath}
\usepackage{longtable}
\usepackage{multirow}
\usepackage{booktabs}
\usepackage{mathtools}
\usepackage{afterpage}
\usepackage{float}

\newcommand{\AGAMA}{\texttt{AGAMA}}
\newcommand{\MWMMXVII}{\texttt{MW2017}}
\newcommand{\HDBSCAN}{\texttt{HDBSCAN}}
\newcommand{\peri}{\text{peri}}
\newcommand{\apo}{\text{apo}}
\newcommand{\maxtext}{\text{max}}

\renewcommand{\arraystretch}{0.5}

\shorttitle{RPE CDTGs}
\shortauthors{Shank et al.}

\begin{document}

\title{The \emph{R}-Process Alliance: Chemo-Dynamically Tagged Groups II.  An Extended Sample of Halo $r$-Process-Enhanced Stars}

\author[0000-0001-9723-6121]{Derek Shank}
\affiliation{Department of Physics and Astronomy, University of Notre Dame, Notre Dame, IN 46556, USA}
\affiliation{Joint Institute for Nuclear Astrophysics -- Center for the Evolution of the Elements (JINA-CEE), USA}

\author[0000-0003-4573-6233]{Timothy C. Beers}
\affiliation{Department of Physics and Astronomy, University of Notre Dame, Notre Dame, IN 46556, USA}
\affiliation{Joint Institute for Nuclear Astrophysics -- Center for the Evolution of the Elements (JINA-CEE), USA}

\author[0000-0003-4479-1265]{Vinicius M. Placco}
\affiliation{NSF’s NOIRLab, 950 N. Cherry Ave., Tucson, AZ 85719, USA}

\author[0000-0003-3246-0290]{Dmitrii Gudin}
\affiliation{Department of Mathematics, University of Maryland, College Park, MD 20742, USA}

\author{Thomas Catapano}
\affiliation{Department of Physics and Astronomy, University of Notre Dame, Notre Dame, IN 46556, USA}

\author[0000-0002-5463-6800]{Erika M.\ Holmbeck}
%\affiliation{Center for Computational Relativity and Gravitation, Rochester Institute of Technology, Rochester, NY 14623, USA} 
\affiliation{Observatories of the Carnegie Institution for Science, 813 Santa Barbara St., Pasadena, CA 91101, USA}
\affiliation{Joint Institute for Nuclear Astrophysics -- Center for the Evolution of the Elements (JINA-CEE), USA}
\affiliation{Hubble Fellow}

\author[0000-0002-8504-8470]{Rana Ezzeddine}
\affiliation{Department of Astronomy, University of Florida, Bryant Space Science Center, Gainesville, FL 32611, USA}
\affiliation{Joint Institute for Nuclear Astrophysics -- Center for the Evolution of the Elements (JINA-CEE), USA}

\author[0000-0001-5107-8930]{Ian U. Roederer}
\affiliation{Department of Astronomy, University of Michigan, 1085 S. University Ave., Ann Arbor, MI 48109, USA}
\affiliation{Joint Institute for Nuclear Astrophysics -- Center for the Evolution of the Elements (JINA-CEE), USA}

\author[0000-0002-5095-4000]{Charli M.\ Sakari}
\affiliation{Department of Physics and Astronomy, San Francisco State University, San Francisco, CA 94132, USA}

\author[0000-0002-2139-7145]{Anna Frebel}
\affiliation{Department of Physics and Kavli Institute for Astrophysics and Space Research, Massachusetts Institute of Technology, Cambridge, MA 02139, USA}
\affiliation{Joint Institute for Nuclear Astrophysics -- Center for the Evolution of the Elements (JINA-CEE), USA}

\author[0000-0001-6154-8983]{Terese T.\ Hansen}
%\affiliation{George P.~and Cynthia Woods Mitchell Institute for Fundamental Physics and Astronomy, Texas A\&M University, College Station, TX 77843, USA}
%\affiliation{Department of Physics and Astronomy, Texas A\&M University, College Station, TX 77843, USA}
\affiliation{Department of Astronomy, Stockholm University, Albanova
University Centre, SE-106 91 Stockholm, Sweden}

\date{\today}

\begin{abstract}
    Orbital characteristics based on Gaia Early Data Release 3 astrometric parameters are analyzed for ${\sim} 1700$ $r$-process-enhanced (RPE; [Eu/Fe] $> +0.3$) metal-poor stars ([Fe/H] $\leq -0.8$) compiled from the $R$-Process Alliance, the GALactic Archaeology with HERMES (GALAH) DR3 survey, and additional literature sources. We find dynamical clusters of these stars based on their orbital energies and cylindrical actions using the \HDBSCAN ~unsupervised learning algorithm. We identify $36$ Chemo-Dynamically Tagged Groups (CDTGs) containing between $5$ and $22$ members; $17$ CDTGs have at least $10$ member stars. Previously known Milky Way (MW) substructures such as Gaia-Sausage-Enceladus, the Splashed Disk, the Metal-Weak Thick Disk, the Helmi Stream, LMS-1 (Wukong), and Thamnos are re-identified. Associations with MW globular clusters are determined for $7$ CDTGs; no recognized MW dwarf galaxy satellites were associated with any of our CDTGs. Previously identified dynamical groups are also associated with our CDTGs, adding structural determination information and possible new identifications. Carbon-Enhanced Metal-Poor RPE (CEMP-$r$) stars are identified among the targets; we assign these to morphological groups in the Yoon-Beers $A$(C)$_{c}$ vs. [Fe/H] Diagram. Our results confirm previous dynamical analyses that showed RPE stars in CDTGs share common chemical histories, influenced by  their birth environments.  
    
\end{abstract}

%Definitely needs changed

\section{Introduction}\label{sec:Introduction}

\begin{deluxetable*}{l  l  l  l}
\tablecaption{Signatures of Metal-Poor Stars \label{tab:MPsignatures}}
\tablehead{\colhead{Signature} & \colhead{Definition} & \colhead{Abbreviation} & \colhead{Source}}
\startdata
      Main $r$-process & $+0.3 < $ [Eu/Fe] $\leq +0.7$, [Ba/Eu] $< 0.0$ & $r$-I & \cite{Holmbeck2020}\\
      Main $r$-process & $+0.7 < $ [Eu/Fe] $\leq +2.0$, [Ba/Eu] $< 0.0$ & $r$-II & \cite{Holmbeck2020}\\
      Main $r$-process & [Eu/Fe] $> +2.0$, [Ba/Eu] $< -0.5$ & $r$-III & \cite{Cain2020}\\
      %Limited $r$-process & [Eu/Fe] $< +0.3$, [Sr/Ba] $> +0.5$, [Sr/Eu] $> 0.0$ & $r_{\rm lim}$ & \cite{Frebel2018}\\
      %$s$-process & [Ba/Fe] $> +1.0$, [Ba/Eu] $> +0.5$, [Ba/Pb] $> -1.5$ & $s$ & \cite{Frebel2018} \\
      %$s$- and $r$-process & $0.0 <$ [Ba/Eu] $< +0.5$, $-1.0 < $ [Ba/Pb] $< -0.5$ & $s + r$ & \cite{Frebel2018} \\
      %$i$-process & $0.0 < $ [La/Eu] $< +0.6$, [Hf/Ir] $\sim +1.0$ & $i$ & \cite{Frebel2018} \\
      Carbon Enhanced & [C/Fe] $> +0.7$ & CEMP & \cite{Aoki2007} \\
      CEMP and RPE & [C/Fe] $> +0.7$, [Eu/Fe] $> +0.3$, [Ba/Eu] < 0.0 & CEMP-$r$ & \cite{Aoki2007} \\
\enddata
\end{deluxetable*}

The rapid neutron-capture process ($r$-process) governs the formation of the heaviest elements in the universe, and accounts for the production of roughly half of the elements beyond iron. A large source of neutrons is required in order to allow neutron-rich isotopes to form far from stability, where they subsequently decay to stable, or long-lived, isotopes all the way up to uranium (U; $Z = 92$). The $r$-process was first formalized by the revolutionary work of \citet{Burbidge1957} and \citet{Cameron1957}, and later Truran and colleagues (e.g., \citealt{Truran1971}; for historical reviews see \citealt{Truran2002} and \citealt{Cowan2021}), who suggested that core-collapse supernovae were the source of $r$-process elements. Candidate sites that produce a sufficient neutron fluence to result in an $r$-process are limited, and have been speculated to be either magnetorotationally jet-driven supernovae (see \citealt{Mosta2018} for a debate on this source), or mergers of either binary neutron stars or a binary neutron star and black hole system, in addition to the already suggested core-collapse supernovae \citep{Lattimer1974,Woosley1994,Winteler2012,Wanajo2014,Nishimura2015,Thielemann2017}. Observations of the kilonova associated with the gravitational wave event GW$170817$ have shown a definitive astrophysical source of heavy elements created by the $r$-process in binary neutron star mergers \citep{Abbott2017a,Abbott2017b,Drout2017,Shappee2017,Tanaka2017,Watson2019}.

The nature of the $r$-process can also be studied through efforts to classify halo stars into chemical groups (see Table~\ref{tab:MPsignatures} for definitions), furthering the statistics of $r$-process abundance patterns. Large-scale efforts to identify $r$-process-enhanced (RPE) stars have been underway since these stars were first recognized by \citet{Sneden1994} (see, e.g., \citealt{Christlieb2004,Barklem2005,Roederer2014b}). With the rarity of these stars limiting the total number of known moderately $r$-process-enhanced ($r$-I) and highly $r$-process-enhanced ($r$-II stars), the $R$-Process Alliance (RPA) was initiated in 2017 with the goal to dramatically increase the total number of identified RPE stars. Through dedicated spectroscopic analysis efforts \citep{Hansen2018,Sakari2018,Ezzeddine2020,Holmbeck2020}, the RPA has already doubled the number of known $r$-II stars (from $65$ to $137$) across the first four data releases \citep{Holmbeck2020}. Additional RPE stars are expected to be identified in the near future, based on ongoing analysis of over a thousand moderately high-resolution, moderate-S/N ``snapshot" spectra obtained by the RPA over the past few years.

The advent of the Gaia satellite mission \citep{GaiaCollaboration2016a} has allowed for precision astrometric parameters (including parallaxes and proper motions) to be collected for over a billion stars, with millions having measured radial velocities (only available for bright sources with $V \lesssim 14$) in Gaia Early Data Release 3 (EDR3; \citealt{GaiaCollaboration2021}). Since $r$-process elements require high-resolution spectra to measure their abundances, accurate radial velocities are often known for RPE stars from such data, even if Gaia does not have this information. These data can be used to reconstruct the orbits of stars once a suitable Galactic potential is chosen. Stars with similar energies and actions, describing the extent of the stellar orbits, can be attributed to the same progenitor satellite or globular cluster which was subsequently accreted into the Milky Way (MW), dispersing the observed RPE stars to their current positions \citep{Helmi1999a}.

\citet{Roederer2018a} employed unsupervised clustering algorithms to group stellar orbital dynamics for RPE stars, an approach that has proven crucial to determine structures in the MW that are not revealed through large-scale statistical sampling methods. These authors were able to determine the orbits for $35$ $r$-II stars. Multiple clustering tools were applied to the orbital energies and actions to identify stars with similar orbital characteristics. This study revealed eight dynamical groupings comprising between two and four stars each. The small dispersion of each group's metallicity was noted, and accounted for by reasoning that each group was associated with a unique accretion event whose stars shared a common chemical history.

\begin{figure*}[t]
    \includegraphics[width=\textwidth]{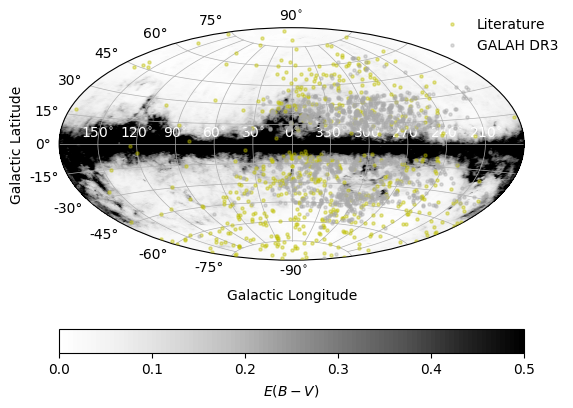}
    \caption{Galactic positions of the RPE Initial Sample of stars, with the literature subset shown as yellow points and the GALAH DR3 subset as gray points. The Galactic reddening map, taken from \citet{Schlegel1998}, and recalibrated by \citet{Schlafly2011}, is shown in the background on a gray scale with darker regions corresponding to larger reddening.}
    \label{fig:galactic_map}
\end{figure*}

\citet{Gudin2021} extended the work by \citet{Roederer2018a}, using a much larger sample of RPE stars, including both $r$-I and $r$-II stars. Utilizing the \HDBSCAN ~algorithm \citep{Campello2013}, $30$ Chemo-Dynamically Tagged Groups (CDTGs)\footnote{The distinction between CDTGS and DTGs is that the original stellar candidates of CDTGs are selected to be chemically peculiar in some fashion, while DTGs are selected from stars without detailed knowledge of their chemistry, other than [Fe/H].} were discovered. Their analysis revealed statistically significant similarities in stellar metallicity, carbon abundance, and $r$-process-element ([Sr/Fe]\footnote{The standard definition for an abundance ratio of an element in a star $(\star)$ compared to the Sun $(\odot)$ is given by $[A/B] = (\log{N_{A}/N_{B}})_{\star} - (\log{N_{A}/N_{B}})_{\odot}$, where $N_{A}$ and $N_{B}$ are the number densities of atoms for elements $A$ and $B$.}, [Ba/Fe], and [Eu/Fe]) abundances within individual CDTGs, strongly suggesting that these stars experienced similar chemical-evolution histories in their progenitor galaxies.

This work aims to expand the efforts of \citet{Roederer2018a} and \citet{Gudin2021}, analyzing the CDTGs present among stars in an updated RPE stellar sample, which includes stars from the literature and the published GALactic Archaeology with HERMES Data Release 3 (GALAH DR3; \citealt{Buder2021}) catalog of metal-poor ([Fe/H] $\leq -0.8$) stars. The procedures employed closely follow the work of \citet{Shank2022a}, which considered DTGs found in the sample of the Best and Brightest selection of \citet{Schlaufman2014} (see \citealt{Placco2019} and \citet{Limberg2021b} for follow-up studies). The association of our identified CDTGs with recognized Galactic substructures, previously known DTGs/CDTGs, globular clusters, and dwarf galaxies is explored, with the most interesting stellar populations being noted for future high-resolution follow-up studies. Statistical analysis of the elemental abundances present in the CDTGs is investigated.
% with a careful inspection of correlations between the elements taken into consideration.

\begin{figure*}[t]
    \includegraphics[width=\textwidth]{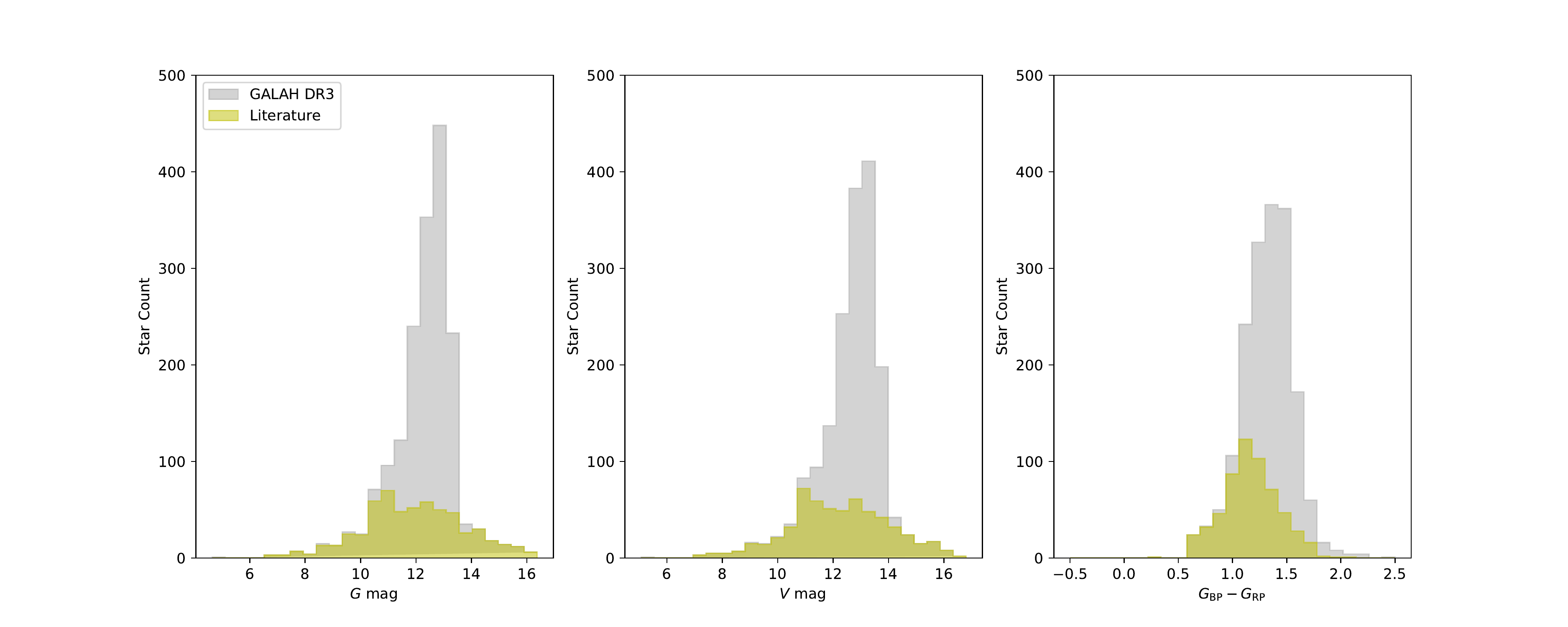}
    \caption{All Panels: The literature subset of the RPE Initial Sample is represented as a yellow histogram; the GALAH DR3 subset is represented as a gray histogram. Left Panel: Histogram of the Gaia $G_{\text{mag}}$ for the RPE Initial Sample. Middle Panel: Histogram of the $V_{\text{mag}}$ for the RPE Initial Sample. Right Panel: Histogram of the Gaia $G_{\text{BP}} - G_{\text{RP}}$ color for the RPE Initial Sample.}
    \label{fig:mag_hist}
\end{figure*}

This paper is outlined as follows. Section \ref{sec:Data} describes the RPE literature and GALAH DR3 sample, along with their associated astrometric parameters and the dynamical parameters. The clustering procedure is outlined in Section \ref{sec:ClusteringProcedure}. Section \ref{sec:StructureAssociations} explores the clusters and their association to known MW structures. %Section~\ref{sec:global_properties} considers the implications for the chemical and dynamical behaviors of the identified CDTGs. 
A statistical analysis of the CDTG abundances is presented in Section~\ref{sec:chemical_structure}. Finally, Section~\ref{sec:Discussion_2} presents a short summary and perspectives on future directions.
\vspace{1.5cm}

\section{Data}\label{sec:Data}

\begin{figure*}[t]
    \includegraphics[width=\textwidth,height=0.8\textheight]{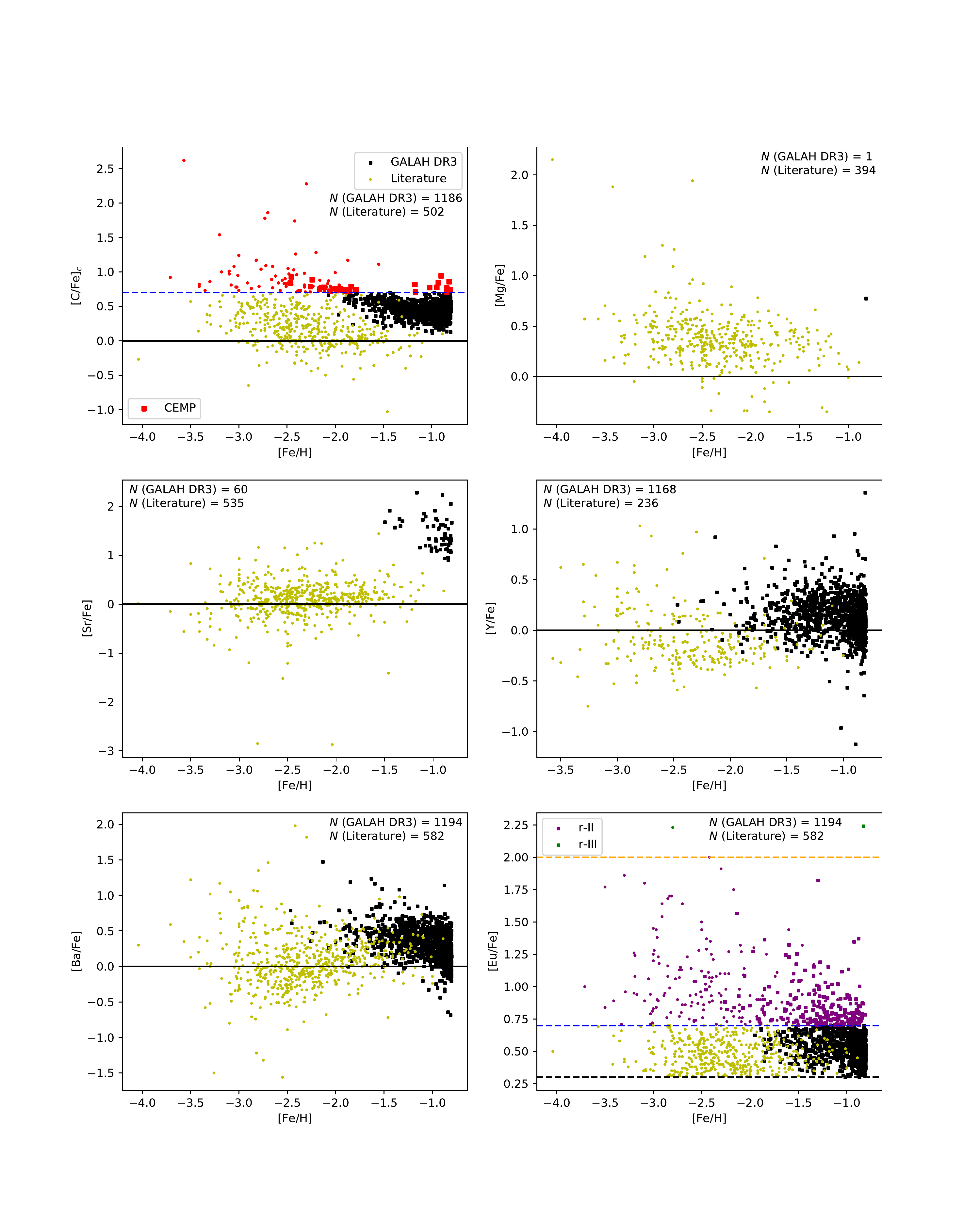}
    \caption{All Panels: The literature subset of the RPE Initial Sample is represented as yellow points; the GALAH DR3 subset is represented as black squares. When there is a classification, such as CEMP, $r$-II, or $r$-III, the points and squares differentiate between literature and GALAH DR3 subsets, respectively. The number of stars with detected abundances in each subset is listed in each panel. Top Left Panel: The corrected carbon abundance ([C/Fe]$_{c}$) of the RPE Initial Sample, corrected from the observed value to account for the depletion of carbon on the giant branch following \citet{Placco2014b}, as a function of metallicity ([Fe/H]). The CEMP cutoff ([C/Fe]$_{c} = +0.7$) is noted as the blue dashed line; CEMP-$r$ stars are indicated with red points for the RPE Initial Sample. The Solar value is indicated as the solid black line. Top Right Panel: The magnesium abundance ([Mg/Fe]) of the RPE Initial Sample, as a function of metallicity ([Fe/H]). The Solar value is indicated with a solid black line. Middle Left Panel: The strontium abundance ratio ([Sr/Fe]) of the RPE Initial Sample, as a function of metallicity ([Fe/H]). The Solar value is indicated with a solid black line. Middle Right Panel: The yttrium abundance ratio ([Y/Fe]) of the RPE Initial Sample, as a function of metallicity ([Fe/H]). The Solar value is indicated with a solid black line. Bottom Left Panel: The barium abundance ratio ([Ba/Fe]) of the RPE Initial Sample, as a function of metallicity ([Fe/H]). The Solar value is indicated with a solid black line. Bottom Right Panel: The europium abundance ratio ([Eu/Fe]) of the RPE Initial Sample, as a function of metallicity ([Fe/H]). The $r$-I cutoff ([Eu/Fe]$ = +0.3$) is noted as the black dashed line;  the $r$-II cutoff ([Eu/Fe]$ = +0.7$) is noted as the blue dashed line, and the $r$-III cutoff ([Eu/Fe]$ = +2.0$) is noted as the orange dashed line. The $r$-I stars are indicated as black (GALAH DR3) and yellow (literature) points, $r$-II stars are indicated as purple points, and $r$-III stars are indicated as green points.}
    \label{fig:abundances}
\end{figure*}

\begin{figure*}[!]
    \includegraphics[width=\textwidth,height=0.87\textheight]{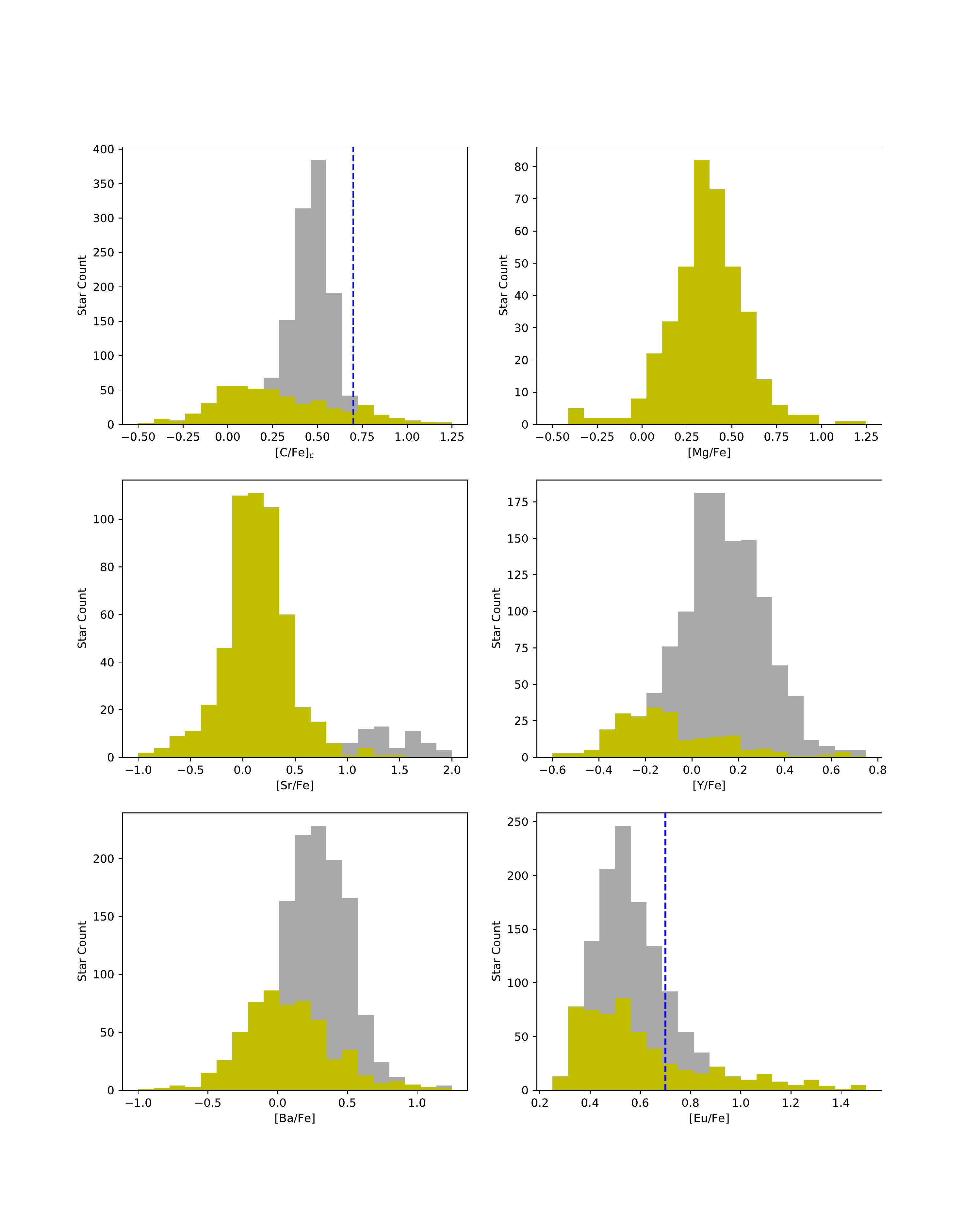}
    \caption{All Panels: The literature subset of the RPE Initial Sample is represented as a yellow histogram; the GALAH DR3 subset is represented as a gray histogram. Note that a few stars have abundances outside the ranges shown. This is to better show the distributions, with outliers being easily identified in Figure~\ref{fig:abundances}. Top Left Panel: Histogram of the corrected carbon abundance ([C/Fe]$_{c}$) of the RPE Initial Sample, corrected from the observed value to account for the depletion of carbon on the giant branch following \citet{Placco2014b}. The CEMP cutoff ([C/Fe]$_{c} = +0.7$) is noted as the blue dashed line. Top Right Panel: Histogram of the magnesium abundance ([Mg/Fe]) of the RPE Initial Sample. Middle Left Panel: Histogram of the strontium abundance ratio ([Sr/Fe]) of the RPE Initial Sample. Middle Right Panel: Histogram of the yttrium abundance ratio ([Y/Fe]) of the RPE Initial Sample. Bottom Left Panel: Histogram of the barium abundance ratio ([Ba/Fe]) of the RPE Initial Sample. Bottom Right Panel: Histogram of the europium abundance ratio ([Eu/Fe]) of the RPE Initial Sample;  the $r$-II cutoff ([Eu/Fe]$ = +0.7$) is noted as the blue dashed line.}
    \label{fig:abundances_histogram}
\end{figure*}

\subsection{Construction of the RPE Initial Sample}\label{subsec:initial_sample}

A literature compilation of RPE stars and known RPE stars in GALAH DR3 form the basis for our data set, as described below.

\subsubsection{The RPE Literature Sample}\label{subsubsec:construct_literature}

A literature search for RPE stars, including the published material from the RPA, is constructed from the most recent version of \href{https://jinabase.pythonanywhere.com/}{JINAbase} \footnote{\href{https://jinabase.pythonanywhere.com/}{https://jinabase.pythonanywhere.com/}.} \citep{Abohalima2018}.  This version crucially includes both the abundances relative to the Sun, as well as the absolute abundances. Unlike the work presented in \citet{Gudin2021}, the literature sample is chosen based on the absolute abundances, and scaled to the Solar atmospheric abundances presented in \citet{Asplund2009}. A restriction on the stellar parameters is applied with [Fe/H] $\leq -0.8$ and $4250 \leq T_{\rm eff} \,\rm{(K)} \leq 7000$ being the target range for RPE stars. The RPE sources are all spectroscopic surveys, and while there is not a uniform methodology in common between the analyses for determining stellar parameters, the methodologies do not differ much in their results (see Fig. 5 of \citealt{Sakari2018}). There are a total of $582$ RPE stars from the literature with [Fe/H] $\leq -0.8$ and $4250 \leq T_{\rm eff} \,\rm{(K)} \leq 7000$ that satisfy the requirements for classification as $r$-I ($426$ stars), $r$-II ($155$ stars), or $r$-III ($1$ star) \citep{McWilliam1995a,McWilliam1995b,Ryan1996,Burris2000,Fulbright2000,Johnson2002,Cohen2004,Honda2004,Aoki2005,Barklem2005,Ivans2006,Preston2006,Franccois2007,Lai2008,Hayek2009,Behara2010,For2010,Roederer2010a,Hollek2011,Hansen2012,Masseron2012,Roederer2012c,Roederer2012a,Aoki2013,Ishigaki2013,Casey2014b,Placco2014a,Roederer2014b,SiqueiraMello2014,Hansen2015,Howes2015,Jacobson2015,Howes2016,Aoki2017,Hill2017,Placco2017,Cain2018,Hansen2018,Hawkins2018,Holmbeck2018,Sakari2018,Mardini2019,Sakari2019,Valentini2019,Xing2019,Cain2020,Bandyopadhyay2020,Ezzeddine2020,Hanke2020,Holmbeck2020,Mardini2020,Placco2020,Rasmussen2020,Yong2021a,Yong2021b,Naidu2022,Roederer2022,Zepeda2022}. Limited-$r$ stars are not discussed in this work and left to future studies.

\subsubsection{The GALAH DR3 Sample}\label{subsubsec:construct_GALAH}

The remainder of our sample is taken from GALAH DR3. The abundances in GALAH DR3 are subject to known quality checks, which are crucial to take into consideration; we only kept stars that have no concerns with their abundance determinations satisfying \texttt{flag\_X\_fe} $= 0$ and \texttt{snr\_c3\_iraf} $> 30$ \citep{Buder2021}. We have employed the same procedure as the literature sample to put the GALAH DR3 stars on the same Solar scale as presented in \citet{Asplund2009}. We restrict stellar values to [Fe/H] $\leq -0.8$ and $4250 \leq T_{\rm eff} \,\rm{(K)} \leq 7000$, the same as the RPE literature subset. We perform this stellar parameter cut to stay consistent with the RPE literature sample, and dynamical studies require a metallicity cut to allow MW substructure formed from accreted dwarf galaxies to be more easily detected \citep{Yuan2020a}. The sample was then cleaned for stars that already were in the RPE Initial Sample Literature subset, though this was only a handful of stars. While the GALAH DR3 sample has spectroscopically derived stellar parameters, there is not sufficient overlap between the stars in the RPE literature sample and the GALAH DR3 sample to comment on the validity for RPE stars. However, Fig. 6 of \citet{Buder2021} shows that the stellar parameters obtained by GALAH DR3 do not differ much from Gaia FGK Benchmark stars. The stellar parameter cut yields $1194$ metal-poor stars from GALAH DR3 that satisfy the requirements for classification as $r$-I ($967$ stars), $r$-II ($226$ stars), or $r$-III ($1$ star).

We henceforth refer to the union of the two above samples as the RPE Initial Sample, and list them in Table~\ref{tab:initial_data_descript} in the Appendix.  In the print edition, only the table description is provided; the full table is available only in electronic form.

\begin{figure}[t]
    \includegraphics[width=0.48\textwidth]{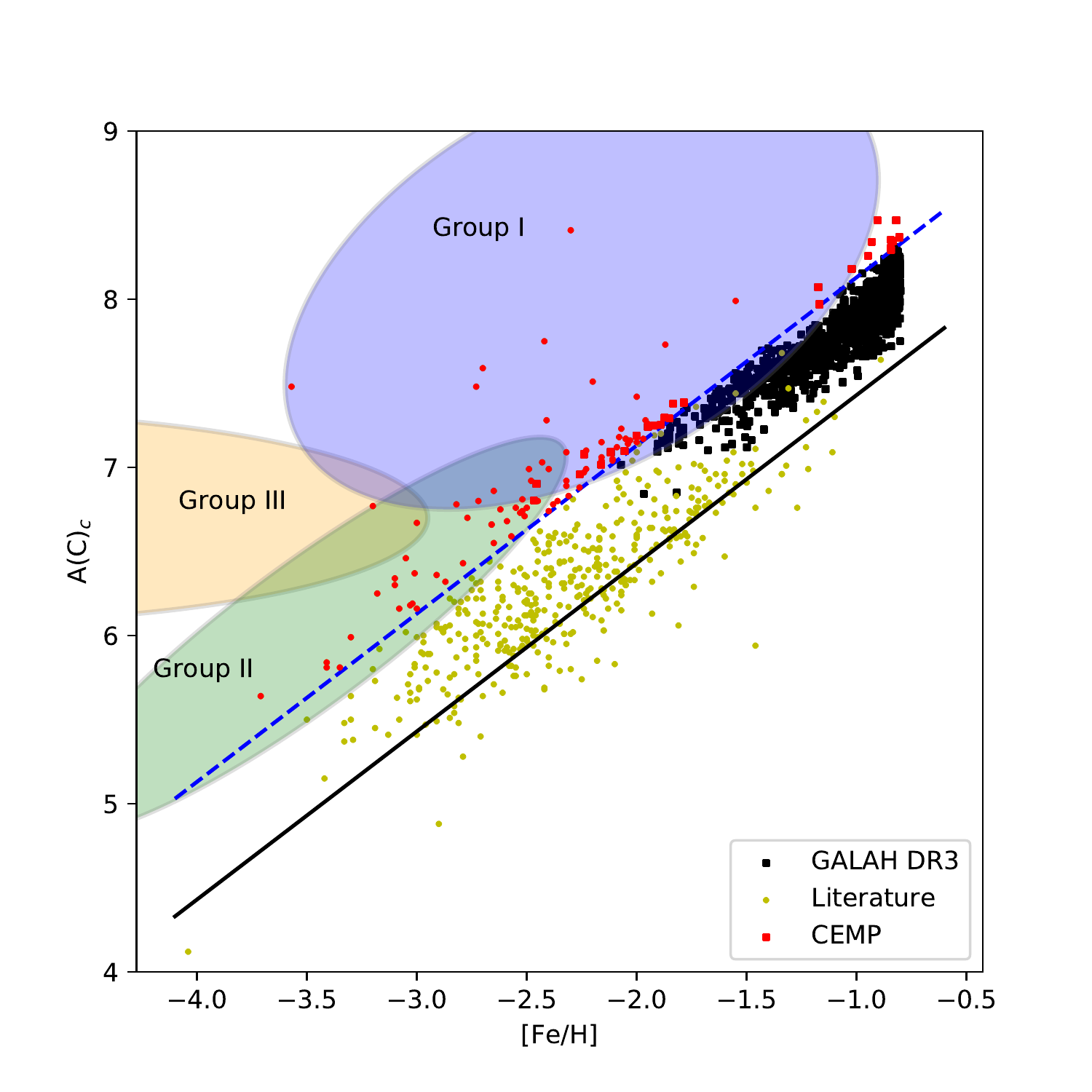}
    \caption{The Yoon-Beers Diagram of $A{\rm (C)}_c$, as a function of [Fe/H], for stars in the RPE Initial Sample, corrected from the observed value to account for the depletion of carbon on the giant branch following \citet{Placco2014b}. The literature subset of the RPE Initial Sample is represented as yellow points; the GALAH DR3 subset is represented as a black squares. For the CEMP classification in red, the points and squares differentiate between literature and GALAH DR3 subsets, respectively. The CEMP cutoff ([C/Fe]$_{c} = +0.7$) is indicated with a blue dashed line. The CEMP-$r$ stars for the RPE Initial Sample are shown as red points. [C/Fe] $= 0$ is indicated with a solid black line. The ellipses represent the three different morphological groups of CEMP stars (See Figure~1 in \citealt{Yoon2016} for a comparison and more information).}
    \label{fig:yoon_beers}
\end{figure}

The stars from the RPE Initial Sample were then cross-matched with Gaia Early Data Release 3 (EDR3; \citealt{GaiaCollaboration2021}), using the same methods outlined in \citet{Shank2022a}.
%a $5 \arcsec$ radius to find their $6$-D astrometric parameters. To validate the match for each star, confirmation was performed by checking that the stellar magnitudes agreed to within $0.5$ mag between the sources. The RPE Sample was mostly taken from \textit{V} magnitudes supplied by the AAVSO Photometric All Sky Survey (APASS; \citealt{Henden2014}) Data Release 9 (DR9; \citealt{Henden2016}), with various other sources listed in the Appendix tables supplying the rest. The corresponding matches were then compared with the \textit{V} magnitude utilizing the transformation from Gaia magnitudes \textit{G}, \textit{G}$_{\text{BP}}$, and \textit{G}$_{\text{RP}}$ in EDR3 provided in Table C.2 of \citet{Riello2021}.
Figure~\ref{fig:galactic_map} shows the spatial distribution of these subsets of RPE stars in Galactic coordinates. 
%There are not a lot of stars in the Northern Hemisphere, as seen to the left of the figure, especially with GALAH being a Southern Hemisphere survey \citep{Buder2021}.
A comparison of the magnitudes and colors for the RPE literature and GALAH DR3 subsets can be seen in Figure~\ref{fig:mag_hist}. From inspection of the figure, it is clear the GALAH DR3 subset of the RPE Initial Sample peaks at fainter magnitudes and redder colors compared to the literature RPE literature subset. RPE stars from the literature are relatively bright, due to the need for high-resolution spectra to detect the $r$-process elements.  Bright stars can be studied with smaller aperture telescopes, and require less observation time on larger aperture telescopes. GALAH DR3 (all spectra obtained with the AAT 3.9-m telescope) includes spectra taken for somewhat fainter stars.

\begin{figure}
    \includegraphics[width=0.5\textwidth]{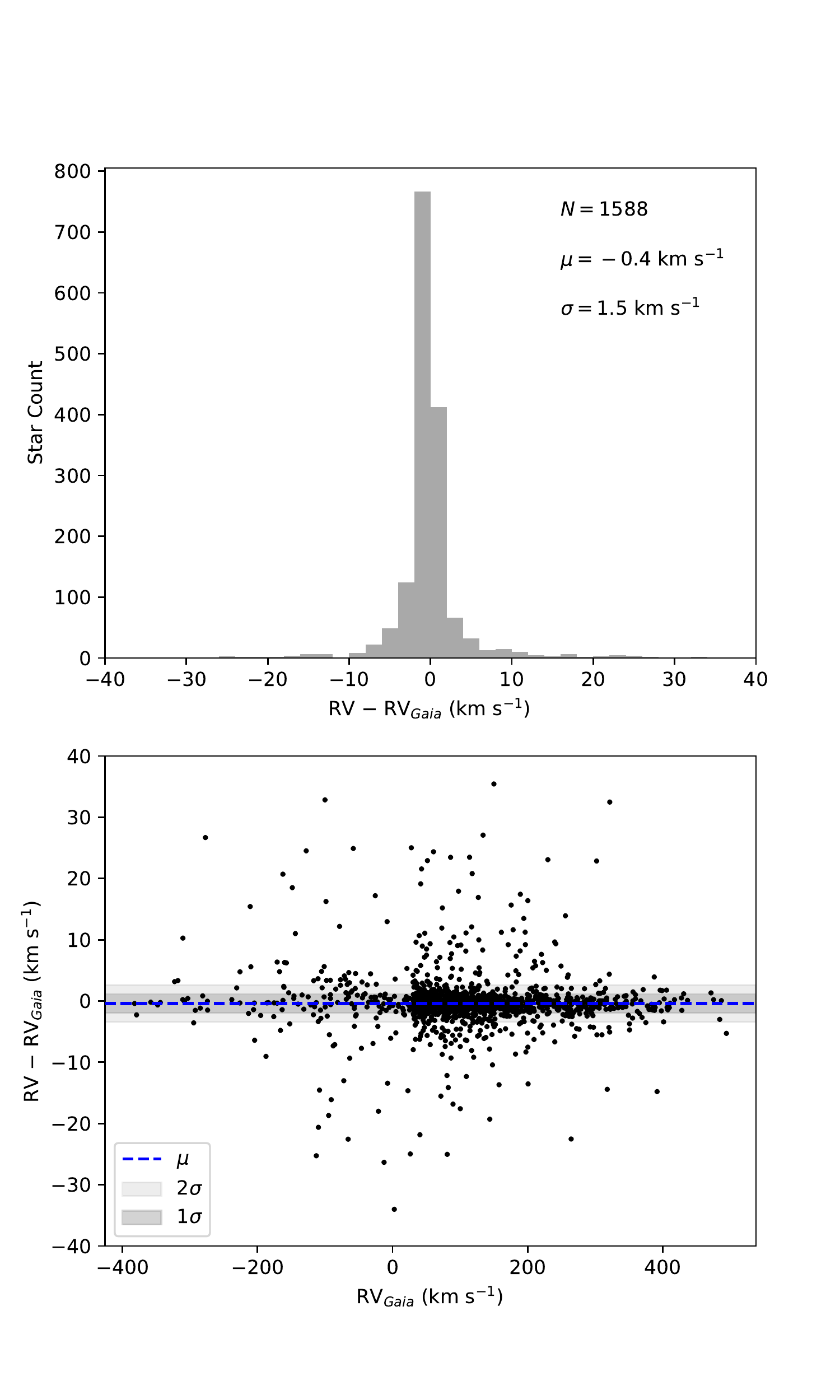}
    \caption{Top Panel: Histogram of the residuals of the difference between the source radial velocities and the Gaia EDR3 values in the RPE Initial Sample. The biweight location and scale are noted. Bottom Panel: The residuals between the source and Gaia EDR3 radial velocities in the RPE Initial Sample, as a function of the Gaia radial velocities. The blue dashed line is the biweight location of the residual difference ($\mu = -0.4$ km s$^{-1}$), while the shaded regions represent the first (1$\sigma = 1.6$ km s$^{-1}$), and second (2$\sigma = 3.2$ km s$^{-1}$) biweight scale ranges.}
    \label{fig:rv_hist}
\end{figure}

\begin{figure}[t]
    \includegraphics[width=0.5\textwidth]{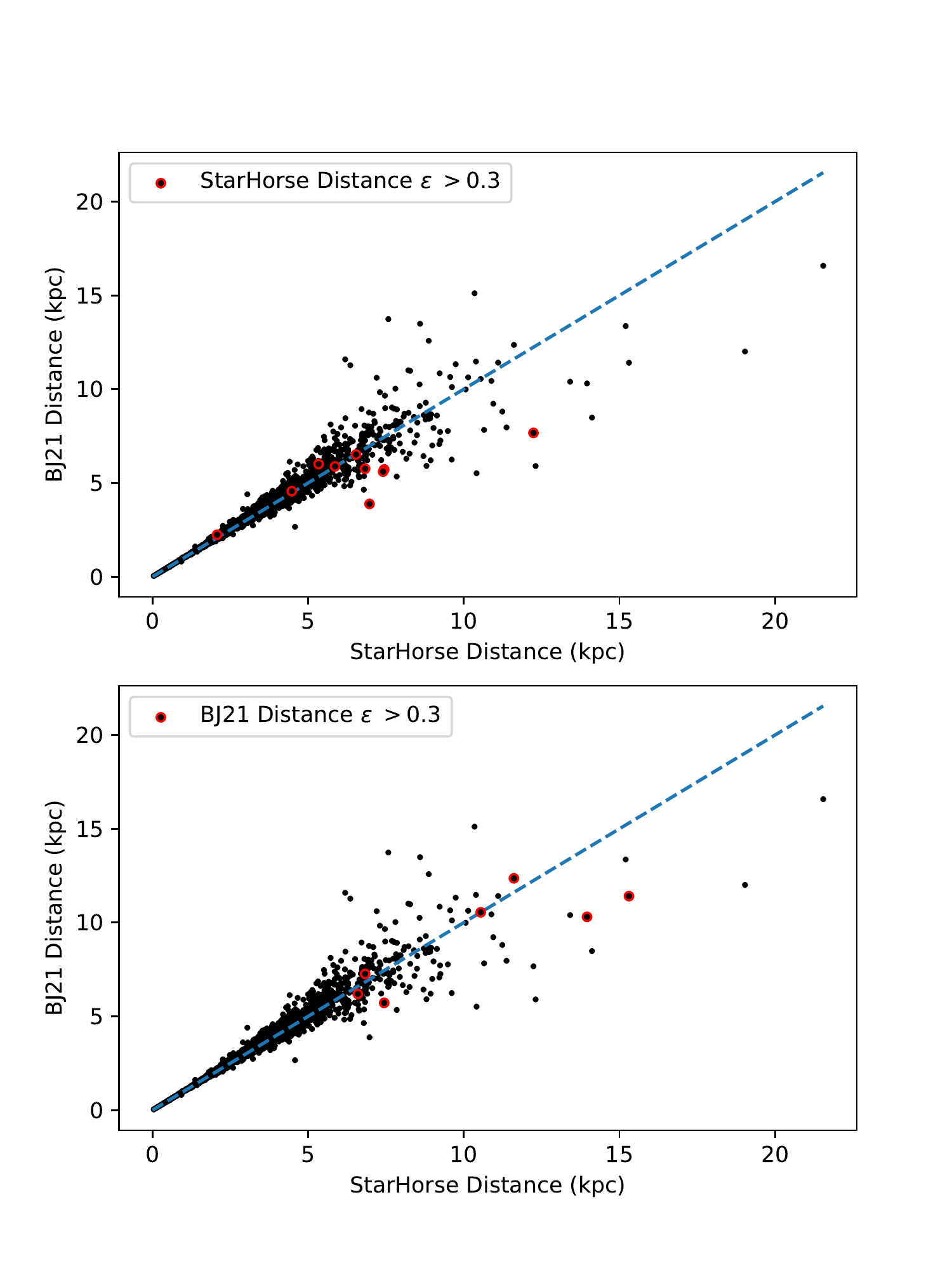}
    \caption{Top Panel: Comparison between the StarHorse distances and BJ21 distances in the RPE Initial Sample. Stars with red outlined points indicate that the relative distance error in StarHorse is $\epsilon > 0.3$. Bottom Panel: The same comparison as the top panel, but with the red outlined points indicating the relative distance error in BJ21 is $\epsilon > 0.3$. In both plots the dashed line indicates a one-to-one comparison between the two samples.}
    \label{fig:dist_comp}
\end{figure}

%\vfill\eject
%\vspace{1cm}

\begin{figure*}
    \includegraphics[width=\textwidth]{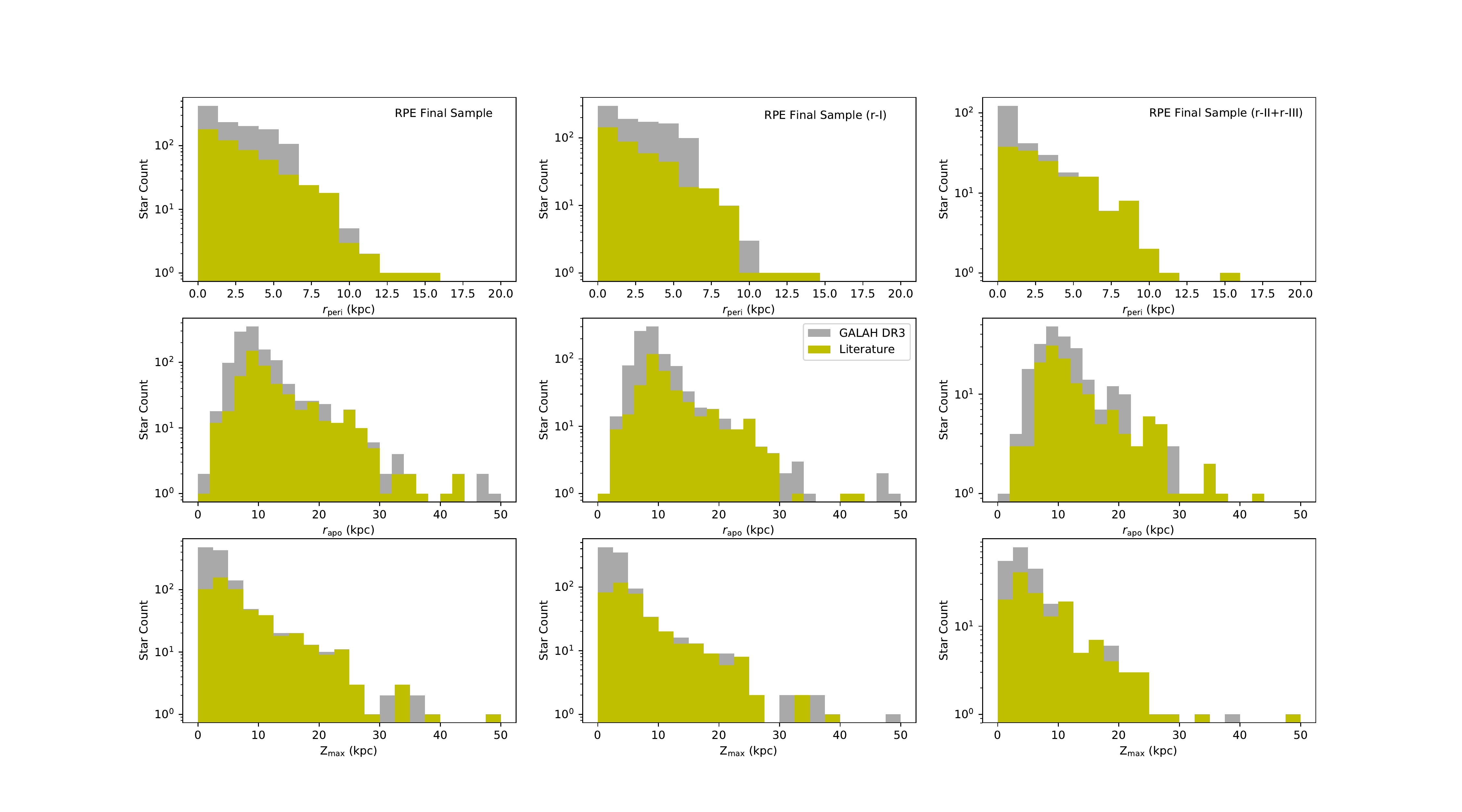}
    \caption{All Panels: The literature subset of the RPE Final Sample is represented as a yellow histogram; the GALAH DR3 subset is represented as a gray histogram. Left Column: Logarithmic histograms of the orbital parameters \textit{r}$_{\peri}$ (top), \textit{r}$_{\apo}$ (middle), and Z$_{\maxtext}$ (bottom) for the RPE Final Sample. Note that a few stars have \textit{r}$_{\apo}$ and Z$_{\maxtext}$ outside the range shown in the panels. Middle Column: Logarithmic histograms of the orbital parameters \textit{r}$_{\peri}$ (top), \textit{r}$_{\apo}$ (middle), and Z$_{\maxtext}$ (bottom) for the RPE Final Sample with only $r$-I stars. Right Column: Logarithmic histograms of the orbital parameters \textit{r}$_{\peri}$ (top), \textit{r}$_{\apo}$ (middle), and Z$_{\maxtext}$ (bottom) for the RPE Final Sample with only $r$-II and $r$-III stars.}
    \label{fig:orb_dist}
\end{figure*}

The available elemental-abundance ratio estimates for stars in the RPE Initial Sample are plotted in Figure~\ref{fig:abundances}, as a function of [Fe/H]. The RPE stars in GALAH DR3 do not offer much in terms of both Mg and Sr, and are included here for the sake of completeness. In the case of Sr, which is a first-peak $r$-process element, Y can be used as a first-peak substitute with more elemental abundances readily available from GALAH DR3; there is no comparable element for Mg. While results using these elements are postulated, future studies will allow further revisions, where necessary, as the information on abundances is updated and expanded. Figure \ref{fig:abundances_histogram} provides histograms of the elemental-abundance ratios considered in the present work. As can be seen from inspection of these figures, stars in the RPE Initial Sample cover a wide range of abundances. This allows the abundance space to be accurately sub-sampled in later stages of our analysis. The Yoon-Beers Diagram of $A(\rm{C})_{c}$ vs. [Fe/H] for these stars is shown in Figure~\ref{fig:yoon_beers}; $A(\rm{C})_{c}$ is the absolute carbon abundance\footnote{The standard definition for absolute abundance of an element X in a star ($\star$) compared to the Sun ($\odot$) is $A(\rm{X}) = \rm{[X/Fe]}_{\star} + \rm{[Fe/H]}_{\star} +\log\epsilon(\rm{X})_{\odot}$.} corrected from the observed value to account for the depletion of carbon on the giant branch, following \citet{Placco2014b}.  For the convenience of later research, the CEMP-$r$ stars are also provided in Table~\ref{tab:cemp} of the Appendix. These stars are included in the analysis, with $28$ GALAH DR3 and $82$ Literature CEMP-$r$ stars. CEMP-$r$ stars are expected to be enriched in their birth clouds by external sources, and as such, do not conflict with the carbon-normal stars that dominate the RPE Initial Sample \citep{Hansen2015}. The RPE Initial Sample has $1393$ $r$-I stars, $381$ $r$-II stars, and $2$ $r$-III stars, for a total of $1776$ RPE stars.

\subsection{Construction of the Final Sample}\label{subsec:rpe_final_sample}

\subsubsection{Radial Velocities, Distances, and Proper Motions}\label{subsec:rv_dist_pm}

Radial velocities, parallaxes, and proper motions for each star are taken from Gaia EDR3, when available. Radial velocities are available for about $90\%$ of the RPE Initial Sample from Gaia EDR3, with typical errors of ${\sim} 1$ km s$^{-1}$. The top panel of Figure~\ref{fig:rv_hist} shows a histogram of the residual differences between the high-resolution radial velocities from the RPE Initial Sample and the Gaia EDR3 values. The biweight location and scale of these differences are $\mu = -0.4$ km s$^{-1}$ and $\sigma = 1.5$ km s$^{-1}$, respectively. The bottom panel of this figure shows the residuals between the high-resolution sources and Gaia EDR3 radial velocities, as a function of the Gaia radial velocities. The blue dashed line is the biweight location, while the shaded regions represent the first (1$\sigma$) and second (2$\sigma$) biweight scale ranges.Gaia EDR3 radial velocities are used, when available, with literature values supplementing when not.

\begin{figure*}[t!]
    \includegraphics[width=\textwidth]{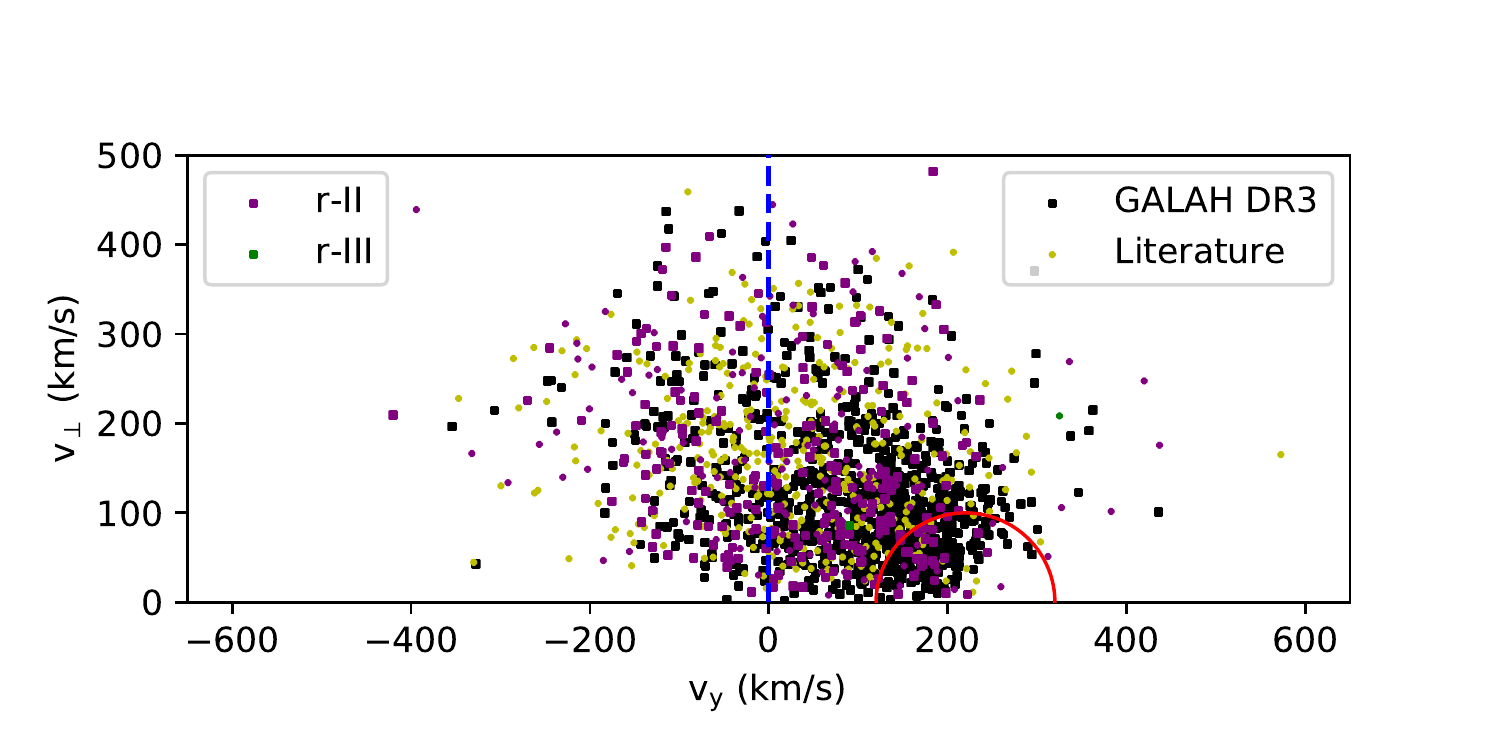}
    \caption{Toomre Diagram of the RPE Final Sample. The literature subset of the RPE Initial Sample is represented as yellow points for $r$-I stars; the GALAH DR3 subset is represented as black squares for $r$-I stars. The $r$-II stars for the RPE Final Sample are indicated in purple, and $r$-III stars are indicated in green. The axes are v$_{\perp} = \sqrt{\text{v}_{x}^{2} + \text{v}_{z}^{2}}$ vs. v$_{y}$. The red circle represents stars within a $100$ km s$^{-1}$ radius from the Local Standard of Rest ($232$ km s$^{-1}$), while the vertical blue dashed line represents the division between prograde (v$_{y} > 0$ km s$^{-1}$) and retrograde (v$_{y} < 0$ km s$^{-1}$) stellar orbits. }
    \label{fig:toomre}
\end{figure*}

The distances to the stars are determined either through the StarHorse distance estimate \citep{Anders2022} or the Bailer-Jones distance estimate (BJ21; \citealt{Bailer-Jones2021}). Parallax values in our RPE Initial Sample from EDR3 have an average error of $\sim 0.02$ mas. The BJ21 and StarHorse distances are determined by a Bayesian approach utilizing the EDR3 parallax, magnitude, and color \citep{Bailer-Jones2021,Anders2022}. The errors are presented for each star in the tables provided in the Appendix. Our adopted distances are chosen following the same procedure outlined in \citet{Shank2022b}, though it is noted here that we prioritize Starhorse distances; the full procedure can be found in Table~\ref{tab:initial_data_descript} in the Appendix.
%We prioritize the StarHorse distances when the relative error (the error divided by the reported value), $\epsilon$, is $ \epsilon < 30\%$. If the StarHorse relative error is $\geq 30\%$, then we adopt the BJ21 distance if the relative error is $ \epsilon < 30\%$. If only one distance estimator is available, then we adopt it. If both distances, or the only available distance, have $\epsilon \geq 30\%$, then we discard the star from the dynamical analysis below. Note that in Figure~\ref{fig:dist_comp} the StarHorse and BJ21 approaches produce similar distances, especially when the distance is smaller than $5$ kpc. 
The proper motions in our RPE Initial Sample from Gaia EDR3 have an average error of $\sim 17$ $\mu$as yr$^{-1}$.

\subsubsection{Dynamical Parameters}\label{subsubsec:DynamicalParameters}

The orbital characteristics of the stars are determined using the Action-based GAlaxy Modelling Architecture\footnote{\url{http://github.com/GalacticDynamics-Oxford/Agama}} (\AGAMA) package \citep{Vasiliev2019}, using the same Solar positions and peculiar motions described in \citet{Shank2022a}\footnote{We adopt a Solar position of ($-8.249$, $0.0$, $0.0$) kpc \citep{GravityCollaboration2020} and Solar peculiar motion in (U,W) as ($11.1$,$7.25$) km s$^{-1}$ \citep{Schonrich2010}, with Galoctocentric Solar azimuthal velocity  \textit{V} $= 250.70$ km s$^{-1}$ determined from \citet{Reid2020}.}, along with the same gravitational potential, \MWMMXVII~ \citep{McMillan2017}. The $6$D astrometric parameters, determined in Section~\ref{sec:Data}, are run through the orbital integration process in \AGAMA, in the same manner as \citet{Shank2022a}, in order to calculate the orbital energy, cylindrical positions and velocities, angular momentum, cylindrical actions, and eccentricity. See \citet{Shank2022a} for definitions of these orbital parameters, and details of the Monte Carlo error calculation.

%The above procedure obtains the orbital parameters if the astrometric parameters are precisely described by the given values. However, these values have errors associated with them, so a method must be developed to estimate the errors in the orbital parameters. This is accomplished through a Monte Carlo sampling over the errors in the astrometric parameters. The procedure that we employ to determine the orbital errors using Monte Carlo sampling is described in detail in \citet{Shank2022a}. 

The RPE Initial Sample of $1776$ stars was cut to exclude stars that are unbound from the MW ($E > 0$) (J$124753.30$-$390802.0$, J$135735.40$-$303249.0$, and J$175159.80$-$475131.0$). Finally, in order to obtain accurate orbital dynamics, we conservatively remove $53$ stars with radial velocity differences $> 15$ km s$^{-1}$ between the Gaia EDR3 values and the high-resolution source values. Most of these stars are expected to be binaries. 

We also considered a cut to remove stars with [Ba/Eu] $< -0.3$ (rather than [Ba/Eu] $< 0$), in order to more confidently select stars with $r$-process enhancement.  We decided not to proceed with this cut, as ultimately including more stars with a wider range of [Ba/Eu] will only serve to increase the abundance dispersion when randomly sampled. Hence, this is a conservative choice; any stars that are included that are in fact not clearly RPE stars (i.e., they have contributions from, e.g., the $s$-process) will only decrease the significance of our dispersion analysis described below\footnote{An explicit test of this cut indeed resulted in a small increase in the statistical significance of both Ba and Eu (with the exception of Eu for the $r$-I sample), following the procedure described in Section~\ref{sec:chemical_structure}.}. 

Application of the above cuts leaves a total sample of $1720$ stars to perform the following analysis. The dynamical parameters of the stars with orbits determined are listed in Table~\ref{tab:final_data_descript} in the Appendix; we refer to this as the RPE Final Sample. In the print edition, only the table description is provided; the full table is available only in electronic form.

\begin{deluxetable*}{l  r  r  c}
\tabletypesize{\scriptsize}
\tablecaption{Identified CDTGs \label{tab:cluster_summary}}
\tablehead{\colhead{CDTG} & \colhead{$N$ Stars} & \colhead{Confidence} & \colhead{Associations with Previously Identified MW Substructure and Dynamical Groups}}
\startdata
1 & $22$ & $93.1\%$ & DS22a:DTG-15, DS22a:DTG-16\\
2 & $21$ & $100.0\%$ & GSE, IR18:E, DG21:CDTG-1, KH22:DTC-15, DS22b:DTG-7 \\
& & & SL22:3, DS22a:DTG-22, EV21:NGC~4833, DS22b:DTG-27, GL21:DTG-37\\
3 & $18$ & $98.9\%$ & DS22b:DTG-1, DS22b:DTG-43\\
4 & $18$ & $68.7\%$ & GSE, DS22b:DTG-30, DG21:CDTG-13, DG21:CDTG-28, DS22a:DTG-18, GC21:Sausage, KH22:DTC-2\\
5 & $17$ & $87.5\%$ & GSE, AH17:VelHel-6, DG21:CDTG-9, DS22a:DTG-3, SL22:3\\
6 & $16$ & $96.1\%$ & Splashed Disk, DS22b:DTG-62, DG21:CDTG-5, DS22a:DTG-19\\
7 & $16$ & $73.8\%$ & MWTD, KH22:DTC-4, DS22b:DTG-14, DG21:CDTG-7, DS22b:DTG-99, DG21:CDTG-6, KM22:C-3, DS22b:DTG-97\\
\enddata
\tablecomments{We adopt the nomenclature for previously identified DTGs and CDTGs from \cite{Yuan2020b}. For example, IR18:E is first initial then last initial of the first author (IR) \citep{Roederer2018a}, the year the paper was published (18), and after the colon is the group obtained by the authors of the paper (E).}
\tablecomments{We use the following references for associations: AH17: \citet{Helmi2017}, GM17: \citet{Myeong2017}, HK18: \citet{Koppelman2018}, GM18a: \citet{Myeong2018b}, GM18b: \citet{Myeong2018c}, IR18: \citet{Roederer2018a}, HL19: \citet{Li2019}, SF19: \citet{Sestito2019}, ZY19: \citet{Yuan2019}, NB20: \citet{Borsato2020}, HL20: \citet{Li2020}, SM20: \citet{Monty2020}, ZY20a: \citet{Yuan2020b}, ZY20b: \citet{Yuan2020a}, GC21: \citet{Cordoni2021}, DG21: \citet{Gudin2021}, CK21: \citet{Kielty2021}, GL21: \citet{Limberg2021a}, KH22: \citet{Hattori2022}, KM22: \citet{Malhan2022}, DS22a: \citet{Shank2022b}, DS22b: \citet{Shank2022a}, and SL22: \citet{SofieLovdal2022}.}\end{deluxetable*}

\begin{deluxetable*}{l  c  c  c  c  c  c  c  c}
\tablecaption{CDTGs Identified by \HDBSCAN \label{tab:cluster_results_stub}}
\tablehead{\colhead{NAME} & \colhead{[Fe/H]} & \colhead{[C/Fe]$_{c}$} & \colhead{[Mg/Fe]} & \colhead{[Sr/Fe]} & \colhead{[Y/Fe]} & \colhead{[Ba/Fe]} & \colhead{[Eu/Fe]}}
\startdata
\multicolumn{8}{c}{CDTG-$1$} \\
\multicolumn{8}{c}{Structure: Unassigned Structure} \\
\multicolumn{8}{c}{Group Assoc: DTG-15: \citet{Shank2022a}} \\
\multicolumn{8}{c}{Group Assoc: DTG-16: \citet{Shank2022a}} \\
\multicolumn{8}{c}{Stellar Assoc: No Stellar Associations} \\
\multicolumn{8}{c}{Globular Assoc: No Globular Associations} \\
\multicolumn{8}{c}{Dwarf Galaxy Assoc: No Dwarf Galaxy Associations} \\
140609001101322 & $-1.15$ & $+0.47$ & $\dots$ & $\dots$ & $+0.21$ & $+0.63$ & $+0.75$ \\
\enddata
\tablecomments{$\boldsymbol{\mu}$ and $\boldsymbol{\sigma}$ represent the biweight estimates of the location and scale for the abundances in the CDTG.}\tablecomments{This table is a stub; the full table is available in the electronic edition.}\end{deluxetable*}

Figure~\ref{fig:orb_dist} provides histograms of \textit{r}$_{\apo}$ (top), \textit{r}$_{\peri}$ (middle), and Z$_{\maxtext}$ (bottom) for stars in the RPE Final Sample. The full sample is shown in the left column, $r$-I stars are shown in the middle column, and $r$-II and $r$-III stars are shown in the right column.  From inspection of this figure, it is clear that the majority of the stars in this sample occupy orbits that take them within the inner-halo region (up to around $15$ to $20$ kpc from the Galactic center), but they also explore regions well into the outer-halo region, up to $\sim 50$ kpc away from the Galactic plane. Although they appear rather similar, according to a Kolmogorov-Smirnov two-sample test between the $r$-I (middle column) and $r$-II plus $r$-III (right column) distributions, the hypothesis that these two samples are drawn from the same parent population is rejected ($p \ll 0.001$) for each of \textit{r}$_{\apo}$ (top), \textit{r}$_{\peri}$ (middle), and Z$_{\maxtext}$ (bottom). It may be  the case that this is due to the different masses of the dwarf satellites in which the $r$-I and $r$-II stars were formed (the majority of RPE stars are likely formed in dwarf satellites, but not all are required to be), but we choose not to speculate further on this point at present.  

Figure~\ref{fig:toomre} is the Toomre Diagram for the RPE Final Sample. The red solid semi-circle indicates whether a stellar orbit is disk-like (inside) or halo-like (outside). There are $234$ ($17\%$ of) $r$-I stars and $30$ ($8\%$ of) $r$-II stars that have disk-like kinematics. Of the disk-like $r$-II stars, there are $22$ that are part of the GALAH DR3 subset of the RPE Final Sample, while the remaining eight are from the literature subset. This difference can be attributed to RPE stars being targeted more in the halo where they have a higher likelihood of detection; most known RPE stars reside in the halo ($85\%$). The blue vertical dashed line indicates whether a stellar orbit is prograde (v$_{y} > 0$ km s$^{-1}$) or retrograde (v$_{y} < 0$ km s$^{-1}$). There are $1026$ ($76\%$) $r$-I stars and $215$ ($64\%$ of) $r$-II stars with prograde orbits. The \citet{Gudin2021} literature sample found an almost fifty-fifty split between RPE stars that have prograde orbits compared to retrograde orbits, while the expanded RPE Final Sample presented here has slightly more prograde stars ($1241$ or $0.70\%$) compared to retrograde stars. Note that simulations performed by \citet{Hirai2022} show that $r$-II stars are slightly favored to be prograde as well. The RPE Final Sample has $1346$ $r$-I stars, $372$ $r$-II stars, and $2$ $r$-III stars, for a total of $1720$ RPE stars.

\vspace{1.0cm}
\section{Clustering Procedure}\label{sec:ClusteringProcedure}

\citet{Helmi2000} were among the first to suggest the use of integrals of motion, in their case orbital energies and angular momenta, to find substructure in the MW using the precision measurements of next-generation surveys that were planned at the time. \citet{McMillan2008} considered the use of actions as a complement to the previously suggested orbital energy and angular momenta, with only the vertical angular momentum being invariant in an axisymmetric potential. Most recently, many authors have employed the orbital energies and cylindrical actions (E,J$_{r}$,J$_{\phi}$,J$_{z}$) to determine the substructure of the MW using Gaia measurements \citep{Helmi2017,Myeong2018b,Myeong2018c,Roederer2018a,Naidu2020,Yuan2020a,Yuan2020b,Gudin2021,Limberg2021a,Shank2022a}.

As described in \citet{Shank2022a}, we employ \HDBSCAN ~in order to perform a cluster analysis over the orbital energies and cylindrical actions from the RPE Sample obtained through the procedure outlined in Section \ref{sec:Data}. The \HDBSCAN ~algorithm\footnote{For a detailed description of the \HDBSCAN ~algorithm visit: \url{https://hdbscan.readthedocs.io/en/latest/how_hdbscan_works.html}} operates through a series of calculations that are able to separate the background noise from denser clumps of data in the dynamical parameters. We utilize the following parameters described in \citet{Shank2022a}: \verb ~min_cluster_size ~$= 5$, \verb ~min_samples ~$= 5$, \verb ~cluster_selection_method ~$=$ \verb 'leaf' , \verb ~prediction_data ~$=$ \verb ~True ~, Monte Carlo samples at $1000$, and minimum confidence set to $20\%$.

Table~\ref{tab:cluster_summary} provides a listing of the $36$ Chemo-Dynamically Tagged Groups\footnote{Chemo-Dynamically Tagged Groups are derived from purely dynamical parameters. The chemical information comes from the RPE selection criteria, thus distinguishing them from Dynamically Tagged Groups (DTGs).} (CDTGs) identified by this procedure, along with their numbers of member stars, confidence values (calculated as described in \citealt{Shank2022a}), and associations described below.  Note that, although a minimum confidence value of $20\%$ was employed, the actual minimum value found for these CDTGs is $43.0\%$ (CDTG-15); only two other CDTGs have an assigned confidence value less than $50.0\%$ (CDTG-18 and CDTG-35). The average confidence level of the $36$ CDTGs is quite high, at $81.9\%$. In total, there were $375$ stars ($22\%$ of the Final Sample) assigned to the $36$ CDTGs.

The previously known groups are identified using the nomenclature introduced by \citet{Yuan2020b}. For example, IR18:E refers to the first initial then the last initial of the lead author (IR) \citep{Roederer2018a}, the year the paper was published (18), and after the colon is the group name provided by the authors of the paper (E). We use the following references for associations: AH17: \citet{Helmi2017}, GM17: \citet{Myeong2017}, HK18: \citet{Koppelman2018}, GM18a: \citet{Myeong2018b}, GM18b: \citet{Myeong2018c}, IR18: \citet{Roederer2018a}, HL19: \citet{Li2019}, SF19: \citet{Sestito2019}, ZY19: \citet{Yuan2019}, NB20: \citet{Borsato2020}, HL20: \citet{Li2020}, SM20: \citet{Monty2020}, ZY20a: \citet{Yuan2020b}, ZY20b: \citet{Yuan2020a}, GC21: \citet{Cordoni2021}, DG21: \citet{Gudin2021}, CK21: \citet{Kielty2021}, GL21: \citet{Limberg2021a}, KH22: \citet{Hattori2022}, KM22: \citet{Malhan2022}, DS22a: \citet{Shank2022b}, DS22b: \citet{Shank2022a}, and SL22: \citet{SofieLovdal2022}.

Table~\ref{tab:cluster_results_stub} lists the stellar members of the identified CDTGs, along with their values of [Fe/H], [C/Fe]$_c$, [Mg/Fe], [Sr/Fe], [Y/Fe], [Ba/Fe], and [Eu/Fe], where available.  The last line (in bold font) in the listing for each CDTG provides the mean and dispersion (both using biweight estimates) for each quantity.  Note that for dynamical groups in which only fewer than three measurements of a given element are provided, we list the mean, and code the dispersion as a missing value.

%Table~\ref{tab:cluster_results_element_stub} lists the identified CDTGs where at least half of the member stars have [C/Fe] and [$\alpha$/Fe] detected. 

Table~\ref{tab:cluster_orbit} lists the derived dynamical parameters (and their errors) derived by \AGAMA ~used in our analysis of the identified CDTGs.

\section{Structure Associations}\label{sec:StructureAssociations}

Associations between the newly identified CDTGs are now sought between known MW structures, including large-scale substructures\footnote{Here, the term ``large-scale substructure'' is used to distinguish large over-densities of stars determined in the integral of motion space in the Galaxy, e.g., the substructures presented in \citet{Naidu2020}.}, previously identified dynamical groups, stellar associations, globular clusters, and dwarf galaxies.

\vspace{5cm}

\startlongtable
\begin{deluxetable*}{l  r  c  c  r  r}
\tablecaption{CDTG Dynamical Parameters Determined by \AGAMA \label{tab:cluster_orbit}}
\tablehead{\colhead{Cluster} & \colhead{$N$ Stars} & \colhead{($\langle$v$_{\text{r}}\rangle$,$\langle$v$_{\phi}\rangle$,$\langle$v$_{\text{z}}\rangle$)} & \colhead{($\langle$J$_{\text{r}}\rangle$,$\langle$J$_{\phi}\rangle$,$\langle$J$_{\text{z}}\rangle$)} & \colhead{$\langle$E$\rangle$} & \colhead{$\langle$ecc$\rangle$}\\
\colhead{} & \colhead{} & \colhead{($\sigma_{\langle\text{v}_{\text{r}}\rangle}$,$\sigma_{\langle\text{v}_{\phi}\rangle}$,$\sigma_{\langle\text{v}_{\text{z}}\rangle}$)} & \colhead{($\sigma_{\langle\text{J}_{\text{r}}\rangle}$,$\sigma_{\langle\text{J}_{\phi}\rangle}$,$\sigma_{\langle\text{J}_{\text{z}}\rangle}$)} & \colhead{$\sigma_{\langle\text{E}\rangle}$} & \colhead{$\sigma_{\langle\text{ecc}\rangle}$}\\
\colhead{} & \colhead{} & \colhead{(km s$^{-1}$)} & \colhead{(kpc km s$^{-1}$)} & \colhead{(10$^{5}$ km$^{2}$ s$^{-2}$)} & \colhead{}}
\startdata
CDTG-$1$ & $22$ & ($-48.4$,$86.3$,$-0.9$) & ($278.9$,$535.9$,$66.5$) & $-1.858$ & $0.639$ \\
 & & ($67.6$,$16.6$,$46.0$) & ($57.6$,$106.8$,$24.4$) & $0.028$ & $0.076$ \\
CDTG-$2$ & $21$ & ($29.7$,$23.3$,$-5.5$) & ($567.6$,$174.9$,$54.2$) & $-1.819$ & $0.886$ \\
 & & ($79.1$,$12.7$,$53.6$) & ($34.4$,$94.6$,$26.0$) & $0.023$ & $0.047$ \\
CDTG-$3$ & $18$ & ($-34.6$,$120.4$,$-3.6$) & ($173.5$,$823.0$,$34.3$) & $-1.831$ & $0.473$ \\
 & & ($52.6$,$22.3$,$38.8$) & ($52.7$,$80.1$,$9.9$) & $0.018$ & $0.076$ \\
CDTG-$4$ & $18$ & ($-39.4$,$-8.4$,$-11.2$) & ($794.4$,$-79.5$,$91.0$) & $-1.677$ & $0.937$ \\
 & & ($150.4$,$17.5$,$55.8$) & ($47.8$,$144.9$,$27.3$) & $0.019$ & $0.037$ \\
\enddata
\end{deluxetable*}

\subsection{Milky Way Substructures}\label{subsec:MWSubstructure}

\begin{figure*}[t]
    \centering
    \includegraphics[width=0.98\textwidth,height=0.98\textheight,keepaspectratio]{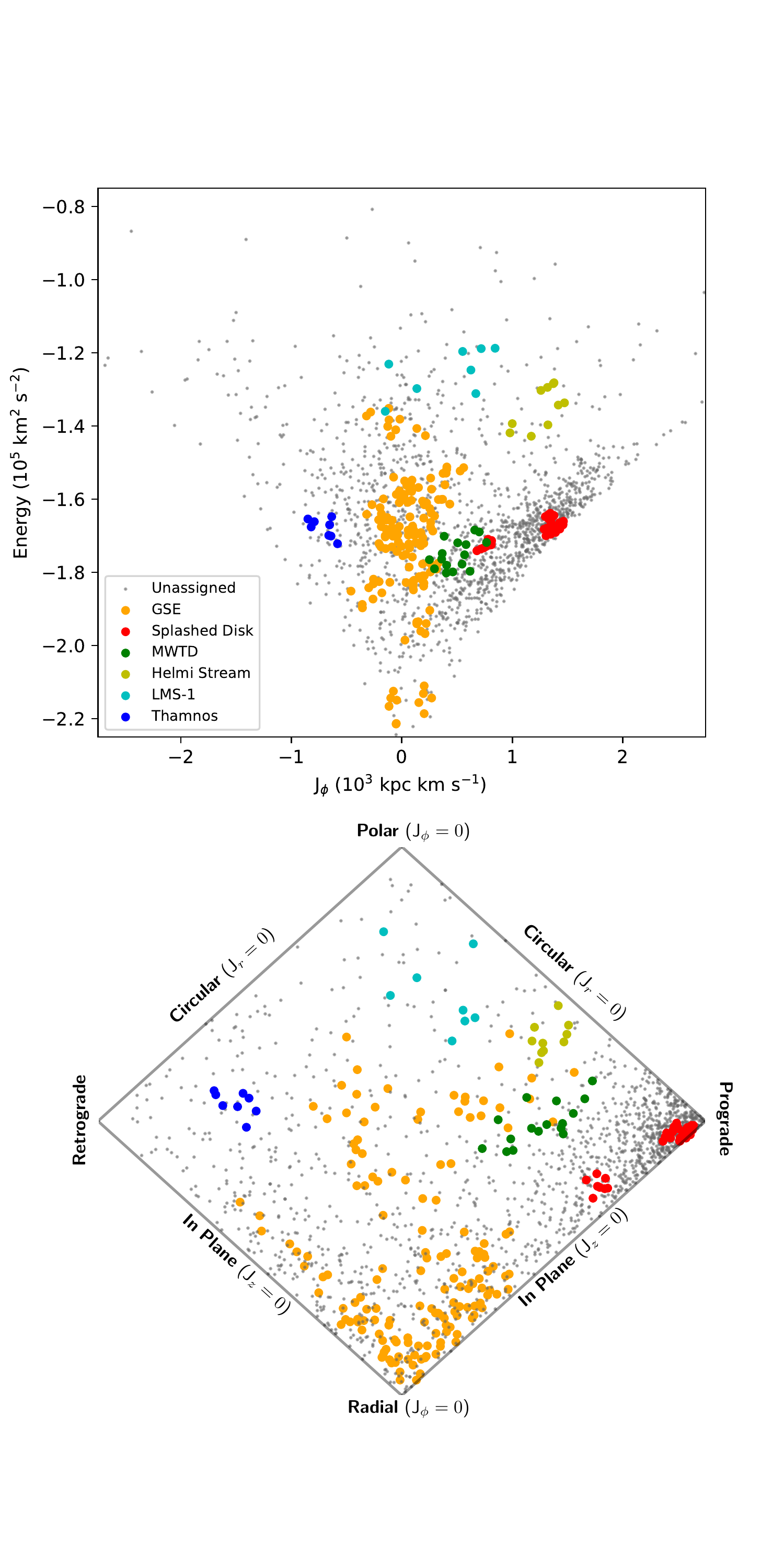}
    \caption{Top Panel: Lindblad Diagram of the identified MW substructures. The different structures are associated with the colors outlined in the legend. Bottom Panel: The projected-action plot of the same substructures. This space is represented by J$_{\phi}$/J$_{\text{Tot}}$ for the horizontal axis and (J$_{\text{z}}$ - J$_{\text{r}}$)/J$_{\text{Tot}}$ for the vertical axis with J$_{\text{Tot}}$ = J$_{\text{r}}$ + $|$J$_{\phi}|$ + J$_{\text{z}}$. For more details on the projected-action space, see Figure~3.25 in \citet{Binney2008}.}
    \label{fig:energy_actions}
\end{figure*}

Analyzing the orbital energies and actions is insufficient to determine separate large-scale substructures. Information on the elemental abundances is crucial due to the differing star-formation histories of the structures, which can vary in both mass and formation redshift \citep{Naidu2020}. The outline for the prescription used to determine the structural associations with our CDTGs is described in \citet{Naidu2020}, and explained in detail in \citet{Shank2022a}. Simple selections are performed based on physically motivated choices for each substructure, excluding previously defined substructures, as the process iterates to decrease contamination between substructures. Following their procedures, we find six predominant MW substructures associated with our CDTGs, listed in Table~\ref{tab:substructures}. This table provides the numbers of stars, the mean and dispersion of their chemical abundances, and the mean and dispersion of their dynamical parameters for each substructure.  The Lindblad Diagram and projected-action plot for these substructures is shown in Figure~\ref{fig:energy_actions}.

\begin{deluxetable*}{l  r  r  r  r  r  r  r  r  c  c  c  r}
\tabletypesize{\tiny}
\tablecaption{Associations with Previously Known MW Substructures \label{tab:substructures}}
\tablehead{\colhead{MW Substructure} & \colhead{$N$ Stars} & \colhead{$\langle$[Fe/H]$\rangle$} & \colhead{$\langle$[C/Fe]$_{\textit{c}}\rangle$} & \colhead{$\langle$[Mg/Fe]$\rangle$} & \colhead{$\langle$[Sr/Fe]$\rangle$} & \colhead{$\langle$[Y/Fe]$\rangle$} & \colhead{$\langle$[Ba/Fe]$\rangle$} & \colhead{$\langle$[Eu/Fe]$\rangle$} & \colhead{($\langle$v$_{\text{r}}\rangle$,$\langle$v$_{\phi}\rangle$,$\langle$v$_{\text{z}}\rangle$)} & \colhead{($\langle$J$_{\text{r}}\rangle$,$\langle$J$_{\phi}\rangle$,$\langle$J$_{\text{z}}\rangle$)} & \colhead{$\langle$E$\rangle$} & \colhead{$\langle$ecc$\rangle$}\\
\colhead{} & \colhead{} & \colhead{} & \colhead{} & \colhead{} & \colhead{} & \colhead{} & \colhead{} & \colhead{} & \colhead{($\sigma_{\langle\text{v}_{\text{r}}\rangle}$,$\sigma_{\langle\text{v}_{\phi}\rangle}$,$\sigma_{\langle\text{v}_{\text{z}}\rangle}$)} & \colhead{($\sigma_{\langle\text{J}_{\text{r}}\rangle}$,$\sigma_{\langle\text{J}_{\phi}\rangle}$,$\sigma_{\langle\text{J}_{\text{z}}\rangle}$)} & \colhead{$\sigma_{\langle\text{E}\rangle}$} & \colhead{$\sigma_{\langle\text{ecc}\rangle}$}\\
\colhead{} & \colhead{} & \colhead{} & \colhead{} & \colhead{} & \colhead{} & \colhead{} & \colhead{} & \colhead{} & \colhead{(km s$^{-1}$)} & \colhead{(kpc km s$^{-1}$)} & \colhead{(10$^{5}$ km$^{2}$ s$^{-2}$)} & \colhead{}}
\startdata
GSE & 155 & $-1.53$ & $+0.38$ & $+0.31$ & $+0.23$ & $+0.12$ & $+0.27$ & $+0.60$ & ($-16.9$,$9.9$,$-5.9$) & ($693.3$,$55.8$,$159.2$) & $-1.718$ & $0.878$\\
 &  & $0.61$ & $0.24$ & $0.19$ & $0.46$ & $0.24$ & $0.31$ & $0.19$ & ($149.1$,$35.0$,$78.4$) & ($327.7$,$192.6$,$160.3$) & \textcolor{white}{+}$0.177$ & $0.090$\\
\hline
Splashed Disk & 36 & $-1.12$ & $+0.45$ & $+0.46$ & $+0.60$ & $+0.11$ & $+0.17$ & $+0.54$ & ($-20.0$,$163.2$,$4.4$) & ($149.3$,$1234.6$,$39.7$) & $-1.682$ & $0.361$\\
 &  & $0.46$ & $0.17$ & $0.03$ & $0.68$ & $0.26$ & $0.26$ & $0.21$ & ($67.9$,$41.6$,$39.7$) & ($107.7$,$256.0$,$26.0$) & \textcolor{white}{+}$0.028$ & $0.140$\\
\hline
MWTD & 16 & $-1.93$ & $+0.54$ & $+0.43$ & $-0.01$ & $+0.03$ & $+0.16$ & $+0.59$ & ($11.4$,$83.9$,$-37.8$) & ($288.2$,$492.1$,$282.5$) & $-1.751$ & $0.632$\\
 &  & $0.72$ & $0.19$ & $0.12$ & $1.15$ & $0.18$ & $0.26$ & $0.19$ & ($102.0$,$40.4$,$89.9$) & ($75.6$,$145.4$,$30.9$) & \textcolor{white}{+}$0.039$ & $0.094$\\
\hline
Helmi Stream & 10 & $-1.99$ & $+0.17$ & $+0.34$ & $+0.09$ & $-0.17$ & $+0.05$ & $+0.49$ & ($-30.8$,$169.9$,$-153.6$) & ($286.0$,$1269.0$,$1063.8$) & $-1.348$ & $0.380$\\
 &  & $0.55$ & $0.36$ & $0.08$ & $0.32$ & $0.17$ & $0.29$ & $0.15$ & ($106.9$,$47.6$,$118.3$) & ($126.3$,$160.2$,$75.6$) & \textcolor{white}{+}$0.054$ & $0.081$\\
\enddata
\end{deluxetable*}

\subsubsection{Gaia-Sausage-Enceladus}\label{subsubsec:GSE}

The most populated substructure identified here is Gaia-Sausage-Enceladus (GSE), which contains $155$ member stars across the associated CDTGs. The selection criteria for GSE is $\langle {\rm ecc} \rangle > 0.7$ \citep{Naidu2020}. These CDTGs are distinct chemo-dynamical groups within GSE, as detected by previous authors as well, showing that, as a massive merger, GSE has distinct dynamical groupings within the progenitor satellite \citep{Yuan2020b,Limberg2021a,Gudin2021,SofieLovdal2022}. GSE is thought to be the remnant of an early merger that distributed a significant number of stars throughout the inner halo of the MW \citep{Belokurov2018,Helmi2018}. The action space determined for the member stars exhibits an extended radial component, a null azimuthal component within errors, and a null vertical component within errors. These orbital properties are the product of the high-eccentricity selection of the CDTGs, and agree with previous findings of GSE orbital characteristics when using other selection criteria \citep{Koppelman2018,Myeong2018b,Limberg2021a,Limberg2022}. 

The $\langle$[Fe/H]$\rangle$ of GSE found in our work ($\sim -1.5$) is rather metal poor, consistent with studies of its metallicity in other dynamical groupings, even though our sample contains more metal-rich stars that could have been associated with GSE \citep{Gudin2021,Limberg2021a,Shank2022a}. The stars that form CDTGs in GSE tend to favor the more metal-poor tail of the substructure, which is also seen in previous dynamical analysis. The $\langle$[Mg/Fe]$\rangle$ ($\sim +0.3$) of GSE exhibits a relatively low level, consistent with the low-Mg structure detected by \citet{Hayes2018} and with Mg levels consistent with accreted structures simulated by \citet{Mackereth2019}, and explained through an accretion origin of GSE. %The Mg abundance pattern seen in GSE is due to accretion of older stellar populations, consistent with known element-abundance patterns. 
We also obtain a $\langle$[C/Fe]$_{\textit{c}}\rangle$ ($\sim +0.4$) for GSE; this elevated C level is indicative of being produced in Type II Supernovae, in agreement with the scenario put forth by \citet{Hasselquist2021} for GSE. 

The RPE stars associated with GSE exhibit Eu enhancement on par with other detected MW susbtructures ($\langle$[Eu/Fe]$\rangle \sim +0.6$). Recently, \citet{Matsuno2021} and \citet{Naidu2021} tracked the formation of RPE stars in GSE, finding high levels of Eu present within identified GSE stars, consistent with the work presented here. Finally, we can associate the globular clusters Ryu 879 (RLGC2), IC 1257, NGC 4833, NGC 5986, NGC 6293, and NGC 6402 (M 14) with GSE, based on CDTGs with similar orbital characteristics and stellar associations of these globular clusters (see Sec. \ref{subsec:GCDG} for details). Note in Figure~\ref{fig:energy_actions} how GSE occupies a large region of the Lindblad Diagram, concentrated in the planar and radial portions of the projected-action plot.

\subsubsection{The Splashed Disk}\label{subsubsec:SD}

The second-most populated substructure identified here is the Splashed Disk (SD), which contains $36$ member stars. The SD is thought to be a component of the primordial MW disk that was kinematically heated during the GSE merger event \citep{Helmi2018,DiMatteo2019,Belokurov2020}. The selection criteria for the SD is $\langle$[$\alpha$/Fe]$\rangle > 0.25 - 0.5\times(\langle$[Fe/H]$\rangle + 0.7)$ \citep{Naidu2020}. The mean velocity components of the SD are consistent with a null radial and vertical velocity, while showing a large positive azimuthal velocity consistent with disk-like stars. The mean eccentricity of these stars is most consistent with disk-like orbits. The SD is the most metal-rich substructure identified here ($\langle$[Fe/H]$\rangle \sim -1.1$). The high $\langle$[Mg/Fe]$\rangle$ ($\sim +0.5$) abundances for the SD shows that these stars are old, and they could be the result of a possible merger event, such as the merger between the MW and GSE progenitor, or heated from a primordial system present within the MW at the time of the GSE merger. The $\langle$[C/Fe]$_{\textit{c}}\rangle$ ($\sim +0.5$) abundance for the SD is high, which is consistent with expectation from the high mean magnesium abundances as a tracer of Type II Supernovae in this substructure \citep{Woosley1995,Kobayashi2006}. 

Note that the SD overlaps with the Metal-Weak Thick Disk in the Lindblad Diagram (Figure~\ref{fig:energy_actions}). This is due to the selection criteria only using metallicity and Mg abundances to determine the SD stars \citep{Naidu2020}. Considering the SD is thought to be composed of stars that have been heated due to the GSE merger event, the positions of the SD stars in the Lindblad Diagram shows a relatively large deviation from disk-like orbits, though some seem to have certainly been less kinematically displaced.

\subsubsection{The Metal-Weak Thick Disk}\label{subsubsec:MWTD}

The third-most populated substructure identified here is the Metal-Weak Thick Disk (MWTD), which contains $16$ member stars. The MWTD is thought to have formed from either a merger scenario, possibly related to GSE, or the result of old stars born within the Solar radius migrating out to the Solar position due to tidal instabilities within the MW \citep{Carollo2019}. The selection criteria for the MWTD is $-0.8 <\langle$[Fe/H]$\rangle < -2.5$, $+0.25 < \langle$[$\alpha$/Fe]$\rangle < +0.45$, and $\langle$J$_{\phi}\rangle > 0.5$ \citep{Naidu2020}. The relative lack of RPE stars in the MWTD agrees with simulations performed by \citet{Hirai2022}, which show that the in situ component does not possess large numbers of highly enhanced $r$-II stars ($\langle$[Eu/Fe]$\rangle \sim +0.6$). The non-existent radial and vertical velocity components, as well as the large positive azimuthal velocity component of the MWTD are all consistent with the velocity distribution for the MWTD from \citet{Carollo2019}, even though the [Fe/H] cut in our dynamical analysis included more metal-rich stars than in their sample. The mean eccentricity distribution found within this substructure is also similar to that reported by \citet{Carollo2019}, showing that the MWTD is a distinct component from the canonical thick disk (TD). Recently, both \citet{An2020} and \citet{Dietz2021} have presented evidence that the MWTD is an independent structure from the TD. 

The distribution in $\langle$[Fe/H]$\rangle$ ($\sim -1.9$) and mean velocity space represents a stellar population consistent with the high-Mg population \citep{Hayes2018} ([Fe/H] $\sim -1.3$), with the mean Mg abundance ($\langle$[Mg/Fe]$\rangle \sim +0.4$) being similar within errors ([Mg/Fe] $\sim +0.3$). The $\langle$[C/Fe]$_{\textit{c}}\rangle$ ($\sim +0.5$) abundance for the MWTD exhibits an enhancement in carbon, possibly pointing to a relation with the strongly prograde CEMP structure found in \citet{Dietz2021}, which was attributed to the MWTD population. While this population is not explicitly recovered, there could be an overlap, and future studies will shed more light on this as new abundance information is explored. 

Interestingly, the MWTD does not have many identified RPE stars compared to detections of this substructure in previous works that did not focus solely on RPE stars \citep{Shank2022b,Shank2022a}. This could be due to the primordial MW disk not being enhanced in $r$-process elements, shown for $r$-II star simulations in \citet{Hirai2022}, or a selection effect, with more stars with Mg abundances needing to be identified in the disk of the MW (note their are only $395$ stars with Mg abundances detected, which are needed to determine the MWTD substructure based on the procedure in \citealt{Naidu2020}). If future abundance measurements show that relatively few RPE stars are identified for the MWTD, then the formation scenarios of the primordial disk can be further constrained. Notice in Figure~\ref{fig:energy_actions} how the MWTD occupies a lower energy component of the disk (the gray dots mostly positioned with prograde orbits) in the Lindblad Diagram, along with being in a more extended disk position as well, a selection which is seen in \citet{Naidu2020}.

\subsubsection{The Helmi Stream}\label{subsubsec:Helmi}

\begin{deluxetable*}{l  l  l  c}
\tabletypesize{\scriptsize}
\tablecaption{Associations of Identified CDTGs \label{tab:interesting_substructure}}
\tablehead{\colhead{Structure} & \colhead{Reference} & \colhead{Associations} & \colhead{Identified CDTGs}}
\startdata
\multirow{6}{*}{MW Substructure} & \multirow{6}{*}{\citet{Naidu2020}} & GSE & 2, 4, 5, 8, 11, 12, 15, 19, 20, 21, 24, 32, 33, 34\\ \cline{3-4}
 &  & Splashed Disk & 6, 9, 23\\ \cline{3-4}
 &  & Helmi Stream & 17\\ \cline{3-4}
 &  & LMS-1 & 25\\ \cline{3-4}
 &  & MWTD & 7\\ \cline{3-4}
 &  & Thamnos & 22\\ \cline{1-4}
\multirow{8}{*}{Globular Clusters} & \multirow{8}{*}{\citet{Vasiliev2021}} & Ryu~879~(RLGC~2) & 8, 20\\ \cline{3-4}
 &  & IC~1257 & 15\\ \cline{3-4}
 &  & NGC~362 & 28\\ \cline{3-4}
 &  & NGC~4833 & 2\\ \cline{3-4}
 &  & NGC~5986 & 11\\ \cline{3-4}
 &  & NGC~6293 & 11\\ \cline{3-4}
 &  & NGC~6397 & 36\\ \cline{3-4}
 &  & NGC~6402~(M~14) & 11\\ \cline{1-4}
\enddata
\end{deluxetable*}

The third-least populated substructure identified here is the Helmi Stream (HS), which contains $10$ member stars. The HS is one of the first detected dynamical substructures in the MW using integrals of motions \citep{Chiba1998,Helmi1999a,Chiba2000}. The selection criteria for the HS is $0.75\times10^{3} < \langle$J$_{\phi}\rangle < 1.7\times10^{3}$ and $1.6\times10^{3} < \langle$L$_{\perp}\rangle < 3.2\times10^{3}$, with L$_{\perp} = \sqrt{\text{L}_{x}^{2} + \text{L}_{y}^{2}}$ \citep{Naidu2020}. The HS has a characteristically high vertical velocity, which separates it from other stars that lie in the disk, and can be seen in the sample here. The large uncertainty on the vertical velocity of the HS members corresponds to the positive and negative vertical velocity components of the stream, with the negative vertical velocity population dominating, consistent with the members determined here \citep{Helmi2020}. 

The $\langle$[Fe/H]$\rangle$ of the HS is more metal poor in this sample ($\langle$[Fe/H]$\rangle \sim -2.0$), compared to the known HS members ([Fe/H] $\sim -1.5$; \citealt{Koppelman2019b}). However, \citet{Limberg2021b} recently noted that the metallicity range of HS is more metal poor than previously expected, with stars reaching [Fe/H] $\sim -2.5$, which is consistent with both the study performed by \citet{Roederer2010a} and the results presented here within errors. \citet{Limberg2021b} also considered $r$-process abundances for the HS, showing that [Eu/Fe] ($> +0.3$) is larger over the wide range of metallicities ($-2.5 \lesssim $[Fe/H]$ \lesssim -1.0$), as also found for the stars reported in this work ($\langle$[Eu/Fe]$\rangle$ $\sim +0.5$). Notice in Figure~\ref{fig:energy_actions} how the HS occupies a relatively isolated space in the Lindblad Diagram, thanks to the large vertical velocity of the stars providing the extra energy compared to the other disk stars.

\subsubsection{LMS-1 (Wukong)}\label{subsubsec:LMS1}

The second-least populated substructure identified here is LMS-1, which contains $8$ member stars. LMS-1 was first identified by \citet{Yuan2020b}, and also detected by \citet{Naidu2020}, who called it Wukong. The selection for LMS-1 is $0.2\times10^{3} < \langle$J$_{\phi}\rangle < 1.0\times10^{3}$, $-1.65\times10^{5} < \langle$E$\rangle < -1.2\times10^{5}$, and $\langle$[Fe/H]$\rangle < -1.45$ \citep{Naidu2020}. This structure is similar to GSE in terms of the velocity component, but is characterized by a higher energy along with a more metal-poor population \citep{Naidu2020}, also found for the small number of stars representing LMS-1 in our sample ($\langle$[Fe/H]$\rangle \sim -2.2$). The carbon and magnesium abundances are also high, which indicates an old population ($\langle$[C/Fe]$_{\textit{c}}\rangle \sim +0.4$ and $\langle$[Mg/Fe]$\rangle \sim +0.4$). LMS-1 exhibits low first-peak $r$-process elements for both strontium and yttrium ($\langle$[Sr/Fe]$\rangle \sim +0.2$ and $\langle$[Y/Fe]$\rangle \sim -0.2$). In Figure~\ref{fig:energy_actions} LMS-1 has a higher energy compared to GSE in the Lindblad Diagram.  However, these stars exhibit a lower eccentricity ($\langle$ ecc $\rangle \sim 0.5$) compared to the GSE stars ($\langle$ ecc $\rangle > 0.7$), forming their own distinct substructure, with a more clear division in the projected-action plot. 

\subsubsection{Thamnos}\label{subsubsec:Thamnos}

The Thamnos substructure also contains $8$ member stars. Thamnos was proposed by \citet{Koppelman2019a} as a merger event that populated stars in a retrograde orbit similar to thick-disk stars. The selection criteria for Thamnos is $-1.5\times10^{3} < \langle$J$_{\phi}\rangle < -0.2\times10^{3}$, $-1.8\times10^{5} < \langle$E$\rangle < -1.6\times10^{5}$, and $\langle$[Fe/H]$\rangle < -1.6$ \citep{Naidu2020}. The low energy and strong retrograde rotation suggest that Thamnos merged with the MW long ago \citep{Koppelman2019a}. Here we find a similar low mean orbital energy and strong mean retrograde motion, and we recover as strong a retrograde motion as in \citet{Koppelman2019a}, within errors. The low mean metallicity ($\langle$[Fe/H]$\rangle \sim -2.1$), consistent with the value reported by \citet{Limberg2021a} ($\langle$[Fe/H]$\rangle \sim -2.2$), and the elevated $\langle$[C/Fe]$_{c}\rangle$ ($\sim +0.4$) of these stars also supports the merger being ancient. The $\langle$[Mg/Fe]$\rangle$ ($\sim +0.5$) is high, also suggesting an old population, consistent with \citet{Kordopatis2020} ([Mg/Fe] $\sim +0.3$). As far as we are aware, this study presents the first known $r$-process-element abundances detected in Thamnos, with $\langle$[Eu/Fe]$\rangle \sim +0.5$ and $\langle$[Ba/Fe]$\rangle \sim +0.1$, and shows that this old system was once subjected to multiple $r$-process events, probably before being accreted into the Galaxy. Notice in Figure~\ref{fig:energy_actions} how Thamnos occupies a space that could be described as a retrograde version of disk stars.

\subsection{Associations to Previously Identified Groups and MW Substructure}\label{subsec:Prev_DTGs_Stellar_assoc}

Separately, we can compare the newly identified CDTGs in this work with other dynamical groups identified by previous authors in order to find structures in common.  We take the mean group dynamical properties from the previously identified groups and compare them to the mean and dispersion for the dynamical parameters of our identified CDTGs. Stellar associations of $5 \arcsec$ are also considered, allowing the identification of stars in our sample that belong to previously identified groups. For details on the previous work used in this process, see \citet{Shank2022a}. The resulting dynamical associations between our identified CDTGs and previously identified groups (along with substructure and globular cluster associations, see Section \ref{subsec:GCDG}) are listed in Table~\ref{tab:interesting_substructure}. The previous groups that are associated to the identified CDTGs in this work are listed in Table~\ref{tab:interesting_groups_substructure}. Table~\ref{tab:cluster_results_stub} lists the individual stellar associations for each of our CDTGs. The works that we use for previous groups are AH17: \citet{Helmi2017}, GM17: \citet{Myeong2017}, HK18: \citet{Koppelman2018}, GM18a: \citet{Myeong2018b}, GM18b: \citet{Myeong2018c}, IR18: \citet{Roederer2018a}, HL19: \citet{Li2019}, SF19: \citet{Sestito2019}, ZY19: \citet{Yuan2019}, NB20: \citet{Borsato2020}, HL20: \citet{Li2020}, SM20: \citet{Monty2020}, ZY20a: \citet{Yuan2020b}, ZY20b: \citet{Yuan2020a}, GC21: \citet{Cordoni2021}, DG21: \citet{Gudin2021}, CK21: \citet{Kielty2021}, GL21: \citet{Limberg2021a}, KH22: \citet{Hattori2022}, KM22: \citet{Malhan2022}, DS22a: \citet{Shank2022b}, DS22b: \citet{Shank2022a}, and SL22: \citet{SofieLovdal2022}. This work is distinguished from previous papers, thanks to the increase in RPE stars compared to \citet{Gudin2021}, and preliminary abundance results for other elements such as Mg and Y. Select identified CDTGs related to each of the large-scale substructures described in Section~\ref{subsec:MWSubstructure} (along with one that is not associated to large-scale substructure) are examined in detail below.
\vspace{1cm}

\subsubsection{CDTG-6}\label{subsubsec:CDTG_6}

CDTG-6 is associated to the Splashed Disk \citep{Naidu2020}, and has interesting associations to previously identified groups. There are three stellar associations made between previously identified groups and CDTG-6, with two coming from DG21:CDTG-5\footnote{We adopt the nomenclature for previously identified DTGs and CDTGs from \cite{Yuan2020b}. For example, DG21:CDTG-5 is represented as the first initial then last initial of the first author (DG) \citep{Gudin2021}, followed by the year the paper was published (21), and after the colon is the group obtained by the authors of the paper (CDTG-5).}, and one coming from DS22a:DTG-62 \citep{Gudin2021,Shank2022b}. DG21:CDTG-5 is associated to the MWTD by the authors \citep{Gudin2021}. DS22a:DTG-62 is not associated to any MW substructure by the authors, though it is noted that there were no measured $\alpha$-element abundances for the stars belonging to DS22a:DTG-62 \citep{Shank2022b}. CDTG-6 has three dynamical associations to previously identified groups -- DG21:CDTG-5, DS22a:DTG-62, and DS22b:DTG-19 \citep{Gudin2021,Shank2022b,Shank2022a}. DS22b:DTG-19 is associated to the MWTD by the authors \citep{Shank2022a}. CDTG-6 seems to point towards an association with either the MWTD or the Splashed Disk, but clearly more Mg abundances are needed before definitive claims can be made. While there were three studies \citep{Gudin2021,Shank2022b,Shank2022a} that had associations corresponding to the MWTD, \citet{Shank2022b} had limited $\alpha$-element abundance information.

\subsubsection{CDTG-7}\label{subsubsec:CDTG_7}

CDTG-7 is the only group associated to the MWTD following the procedure in \citet{Naidu2020}. Three stellar associations are made to DG21:CDTG-7, two to KH22:DTC-4, and one each to DG21:CDTG-6, DS22a:DTG-14, and DS22a:DTG-99. DG21:CDTG-7 is associated to GSE by the authors, while DG21:CDTG-6 is not associated to MW substructure \citep{Gudin2021}. KH22:DTC-4 is not associated to any MW substructure, and is associated to IR18:C by the authors \citep{Hattori2022}. DS22a:DTG-14 is associated to the MWTD, with other associations to HL19:GL-1, DS22b:DTG-2, and DG21:CDTG-6 \citep{Shank2022b}. HL19:GL-1 is not associated to MW substructure by the authors, and also associated to AH17:VelHel-7 \citep{Li2019}. DS22b:DTG-2 is associated with the MWTD, and also associated to HL19:GL-1, DG21:CDTG-6, and DG21:CDTG-8, with DG21:CDTG-8 associated to the MWTD by the authors \citep{Shank2022a,Gudin2021}. DS22a:DTG-99 is not associated to any MW substructure, and also associated to HL19:GL-1 by the authors \citep{Shank2022b}. CDTG-7 is also dynamically associated to KM22:C-3 and DS22a:DTG-97 \citep{Malhan2022,Shank2022b}. KM22:C-3 is not associated to MW substructure by the authors, while DS22a:DTG97 is associated to GSE and also HL20:GR-1 \citep{Malhan2022,Shank2022b}. HL20:GR-1 is associated to AH17:VelHel-7, HL19:GL-1, and HL19:GL-2 by the authors, none of which are recovered here \citep{Li2020}. Interestingly, CDTG-7 has a few associations to GSE, but does not have a strong enough eccentricity ($\langle$ ecc $\rangle ~\sim 0.64$) to be determined as GSE according to the procedure by \citet{Naidu2020} ($\langle$ ecc $\rangle > 0.7$). The chemical information relates this more to the MWTD compared to GSE, with both $\langle$[C/Fe]$_{c}\rangle$ ($\sim +0.5$) and $\langle$[Mg/Fe]$\rangle$ ($\sim +0.4$) being more abundant in CDTG-7 compared to the detected abundances in GSE presented here ($\sim +0.4$ and $\sim +0.3$, respectively). In total, there are $7$ associations between CDTG-7 and previous works, with $2$ associations being related to the MWTD and $2$ to GSE by the previous authors, the rest were not associated to large-scale substructure.

\subsubsection{CDTG-8}\label{subsubsec:CDTG_8}

CDTG-8 is associated with GSE \citep{Naidu2020}, and has interesting associations to previously identified groups. Taking a closer look at CDTG-8, we have six stellar associations for CDTG-8 with two in KH22:DTC-15, and a star in each of GL21:DTG-18, DG21:CDTG-18, DS22a:DTG-57, and DS22a:DTG-58 \citep{Limberg2021a,Gudin2021,Shank2022b}. KH22:DTC-15 is associated to Pontus \citep{Malhan2022}, not discussed in this work, by the authors \citep{Hattori2022}. KH22:DTC-15 is associated to IR18:E, IR18:F, IR18:H, and ZY20a:DTG-38 by the authors as well \citep{Hattori2022}. ZY20a:DTG-38 is related to GSE by the authors \citep{Yuan2020b}. GL21:DTG-18 was not associated to GSE by the authors, or any MW substructure; however, it was associated to ZY20a:DTG-33, which was also not associated to any MW substructure by the authors \citep{Limberg2021a,Yuan2020a}. DG21:CDTG-18 was also not assigned to MW substructure by the authors, while DS22a:DTG-57, and DS22a:DTG-58 were associated to GSE \citep{Gudin2021,Shank2022b}. CDTG-8 is dynamically associated with GC21:Sausage, DS22a:DTG-57, and DS22b:DTG-11 which are all associated to GSE by their authors \citep{Cordoni2021,Shank2022b,Shank2022a}. We also recover a globular cluster match with CDTG-8 to EV21:Ryu 879 (RLGC 2), which is typically associated with GSE dynamics \citep{Callingham2022,Shank2022b,Shank2022a}. There are $8$ associations between CDTG-8 and previous groups, with $5$ being associated to GSE, $1$ to Pontus, and the rest not associated to large-scale substructure by the previous authors.

\subsubsection{CDTG-17}\label{subsubsec:CDTG_17}

Only one group is associated to the HS, CDTG-17. We can see that CDTG-17 has five stellar associations with DG21:CDTG-15, four with NB20:H99, four with SL22:60, three with HK18:Green, and one each with GL21:DTG-3 and DS22a:DTG-42 \citep{Koppelman2018,Borsato2020,Gudin2021,Limberg2021a,Shank2022b,SofieLovdal2022}. All of these groups were associated to the HS by their respective authors, with DG21:CDTG-15 also being associated to GL21:DTG-3 and ZY20a:DTG-3, of which we recover the GL21:DTG-3 association, and with ZY20a:DTG-3 being associated as HS members by the authors \citep{Gudin2021,Limberg2021a,Yuan2020a}. CDTG-17 is also dynamically associated to GM18a:S2, GM18b:S2, DG21:CDTG-15, GL21:DTG-3, and DS22a:DTG-42 which are all associated to HS by their authors \citep{Myeong2018b,Myeong2018c,Gudin2021,Limberg2021a,Shank2022b}. There are $8$ associations between CDTG-17 and previous groups, with all $8$ being assigned to the HS by the previous authors.

\subsubsection{CDTG-22}\label{subsubsec:CDTG_22}

CDTG-22 is associated to Thamnos. CDTG-22 has three stellar associations with two belonging in DG21:CDTG-27, which is identified as belonging to Thamnos by the authors \citep{Gudin2021}, and one belonging in KH22:DTC-16, which is associated to IR18:B by the authors \citep{Hattori2022}. On the other hand, CDTG-22 has a dynamical association to GC21:Sequoia, belonging to Sequoia according to the authors \citep{Cordoni2021}. However, Sequoia is a higher energy structure compared to Thamnos \citep{Koppelman2019b}. CDTG-22 also has a dynamical association to KM22:C-3, which is not associated by the authors \citep{Malhan2022}. This is a case where CDTG-22 could be between the two substructures of Thamnos and Sequoia in terms of energy, and more information is required before a definitive conclusion can be made. There are $4$ associations between CDTG-22 and previous groups, with 1 of those being associated with Thamnos, 1 being associated to Sequoia, and the other 2 not associated to large-scale substructure by the previous authors.

\vspace{0.5cm}
\subsubsection{CDTG-25}\label{subsubsec:CDTG_25}

CDTG-25 is associated to LMS-1 (Wukong) through the procedure outlined in \citet{Naidu2020}. There were six stars from previously identified groups matched with CDTG-25 through a $5 \arcsec$ radius search of the CDTG-25 member stars. Three of the stars are in DG21:CDTG-4, and the other three are in DS22a:DTG-67. DG21:CDTG-4 is not assigned \citep{Gudin2021,Shank2022b}, while DS22a:DTG-67 is associated with LMS-1 \citep{Yuan2020b}. DG21:CDTG-4 was associated with GL21:DTG-2 by the authors, where \citet{Limberg2021a} made a tentative association with LMS-1. These associations seem to strengthen their argument for GL21:DTG-2. CDTG-25 is also dynamically associated with GM18b:Cand10, GM18b:Cand11, DG21:CDTG-4, GL21:DTG-2, and DS22a:DTG-67 \citep{Myeong2018c,Gudin2021,Limberg2021a,Shank2022b}. Both GM18b:Cand10 and GM18b:Cand11 were new groups identified by \citep{Myeong2018c}, though we note that LMS-1 was not discovered until two years later by \citet{Yuan2020b}, meaning that LMS-1 was possibly detected through a dynamical search for groups. There are $5$ associations between CDTG-25 and previous groups, with $2$ being associated to LMS-1 and the other $3$ not associated to large-scale substructure by the previous authors.

\startlongtable
\begin{deluxetable*}{l  l  c}
\tablecaption{Associations of Identified CDTGs with Previous Groups \label{tab:interesting_groups_substructure}}
\tablehead{\colhead{Reference} & \colhead{Associations} & \colhead{Identified CDTGs}}
\startdata
\multirow{28}{*}{\citet{Shank2022b}} & DTG-4 & 18, 27\\ \cline{2-3}
 & DTG-1 & 3\\ \cline{2-3}
 & DTG-5 & 10\\ \cline{2-3}
 & DTG-7 & 2\\ \cline{2-3}
 & DTG-8 & 32\\ \cline{2-3}
 & DTG-11 & 30\\ \cline{2-3}
 & DTG-12 & 20\\ \cline{2-3}
 & DTG-14 & 7\\
\enddata
\end{deluxetable*}

\subsubsection{CDTG-36}\label{subsubsec:CDTG_36}

CDTG-36 is not assigned to any MW substructure, and has two stellar associations to DS22a:DTG-55, and one each with DS22b:DTG-9, KH22:DTC-4, and EV21:NGC 6397 \citep{Vasiliev2021,Hattori2022,Shank2022b,Shank2022a}. All of the past groups are not assigned by their authors to any large-scale MW substructure, with EV21:NGC 6397 being a globular cluster. KH22:DTC-4 is associated to IR18:C by the authors \citep{Hattori2022}. Both DS22a:DTG-55 and DS22b:DTG-9 are also associated with NGC 6397 as well. Although we have presented a stellar association, CDTG-36 has $\langle$[Fe/H]$\rangle \sim -0.9$, as opposed to the metallicity of NGC 6397 of [Fe/H] $\sim -2.0$ \citep{Jain2020}. It is interesting to see stellar associations with NGC 6397 when the average metallicity of CDTG-36 does not compare to the metallicity of NGC 6397, and when both DS22a:DTG-55 and DS22b:DTG-9 are very metal poor ($\langle$[Fe/H]$\rangle \sim -2.5$ and $\langle$[Fe/H]$\rangle \sim -2.5$, respectively) \citep{Shank2022b,Shank2022a}. There are $4$ associations between CDTG-36 and previous groups, with all $4$ not being assigned to large-scale substructure by the previous authors, though $2$ are associated to the globular cluster NGC 6397. CDTG-36 has 2 member stars which are very metal poor, while the other three are just metal poor, which may explain the discrepancy in metallicity observed between CDTG-36 and NGC 6397.

\subsection{Globular Clusters and Dwarf Galaxies}\label{subsec:GCDG}

Both globular clusters and dwarf galaxies have been shown to play an important role in the formation of stars that deviate from the usual chemical-abundance trends in the MW \citep{Ji2016a,Myeong2018a}. Globular clusters can also be a good indicator of galaxy-formation history based on their metallicities and orbits \citep{Woody2021}. From the work of \citet{Vasiliev2021}, we can compare the dynamical properties of $170$ globular clusters to those of the CDTGs we identify. The procedure that is employed is the same one used for previously identified groups and stellar associations introduced in Sec. \ref{subsec:Prev_DTGs_Stellar_assoc}. The dynamics for $45$ dwarf galaxies of the MW (excluding the Large Magellanic Cloud, Small Magellanic Cloud, and Sagittarius) are also explored \citep{McConnachie2020,Li2021}. \citet{Shank2022a} contains details of the orbits of the globular clusters and dwarf galaxies. The same procedure used for previously identified groups was then applied to determine whether a CDTG was dynamically associated to the dwarf galaxy. Stellar associations were also determined for both globular clusters and dwarf galaxies in the same manner as previously identified groups. 

The above comparison exercise led to seven of our identified CDTGs being associated to globular clusters. Table~\ref{tab:interesting_substructure} provides a breakdown of which globular clusters are associated with our CDTGs. The CDTGs associated with globular clusters are expected to have formed in chemically similar birth environments; this is mostly supported through the similar chemical properties of the CDTGs. Associations of globular clusters with Galactic substructure have also been made by \citet{Massari2019} and \citet{Callingham2022}.

Ryu 879 (RLGC 2) (CDTG-8 and CDTG-20), IC 1257 (CDTG-15), NGC 6293 (CDTG-11),  and NGC 6402 (M 14) (CDTG-11) are dynamically associated to their respective groups. On the other hand, NGC 362 (CDTG-28), NGC 4833 (CDTG-2), NGC 5986 (CDTG-11), and NGC 6397 (CDTG-36) have stellar associations to their respective groups. Even though the matched stars in these globular clusters would have individually been associated with the globular cluster orbital parameters, the overall CDTG did not possess sufficiently similar orbital characteristics to be associated.

\begin{itemize}

\item Ryu 879 (RLGC 2) has two dynamical CDTG associations that agree with each other in mean metallicity within errors ($\langle$[Fe/H]$\rangle$ $\sim -1.97 \pm 0.8$ for CDTG-$9$ vs. $\langle$[Fe/H]$\rangle$ $\sim -1.05 \pm 0.16$ for CDTG-$20$), compared to the metallicity of Ryu 879 (RLGC 2) of [Fe/H] $-2.1 \pm 0.3$ \citep{Ryu2018}. Both CDTG-9 and CDTG-20 are associated to GSE in this work, agreeing with the association to Galactic substructure of Ryu 879 (RLGC 2) by \citet{Callingham2022}, while \citet{Massari2019} did not analyze Ryu 879 (RLGC 2), since the globular cluster was only recently discovered at the time of the publication. 

\item IC 1257 is associated to GSE in this work, and \citet{Massari2019}, \citet{Callingham2022}, and \citet{Limberg2022} associate IC 1257 to GSE as well.

%\item Gran 1 is associated to the Bulge by \citet{Callingham2022}, though the authors mention this globular cluster could belong to the Kraken substructure, not explored in this work. \citet{Massari2019} do not analyze Gran 1 since the globular cluster was discovered after the time of publication. Gran 1 is unassociated to any MW substructure in this work.

\item NGC 6293 is associated to the Bulge by both \citet{Massari2019} and \citet{Callingham2022}, though this globular cluster is associated to GSE in this work. CDTG-11, which is associated to NGC 6293, interestingly has the lowest bound orbital energy ($\langle$E$\rangle \sim -2.2$), which actually overlaps with the potential energy of the defined bulge region in \citet{Callingham2022}, but here CDTG-11 is assigned to GSE due to the large eccentricity ($\langle$ecc$\rangle \sim 0.75$). It is possible that this group formed in the bulge with intrinsically large eccentricity.

\item NGC 6402 (M 14) is associated to GSE in this work, but not associated in \citet{Massari2019}. \citet{Callingham2022} associate NGC 6402 (M 14) to the Kraken substructure, not explored in this work.

%\item VV CL002 is associated to the Bulge by \citet{Callingham2022}, though the authors mention this globular cluster could belong to the Kraken substructure, not explored in this work. \citet{Massari2019} do not analyze VVV CL002 since the globular cluster was discovered after the time of publication. VVV CL002 is unassociated to any MW substructure in this work. 

%\item NGC 6235 is associated to Thamnos by the identification in this work, but it is associated to GSE by both \citet{Massari2019} and \citet{Callingham2022}.

\item NGC 362 is not associated to MW substructure in this work, but was associated to GSE by \citet{Massari2019}, \citet{Callingham2022}, and \citet{Limberg2022}. CDTG-28, which was associated to NGC 362, does not have a sufficiently high eccentricity ($\langle$ecc$\rangle \sim 0.5$) in this work to be associated to GSE.

\item NGC 4833 is associated to GSE by both the identification in this work and \cite{Massari2019}, but related to the Kraken substructure, not explored in this work, by \citet{Callingham2022}. 

\item NGC 5986 is associated to the Kraken substructure by \citet{Callingham2022}, while not being assigned by \citet{Massari2019} and associated to GSE in this work. 

\item NGC 6397 is not associated to any substructure in this work, but \cite{Massari2019} find an association to the disk, which is not a part of the substructure routine in this work, while \citet{Callingham2022} associate this globular cluster to the Kraken substructure, not explored in this work.

\end{itemize}

We did not identify any associations of CDTGs to the sample of (surviving) MW dwarf galaxies, either through stellar associations, or through the dynamical association procedure described above. Note that we excluded known RPE stars that are members of recognized dwarf galaxies during the assembly of our field-star sample. As dwarf galaxy astrometric parameters themselves continue to improve, evidence may arise that shows some stripped stars after the dwarf has made a pass near the inner MW, but that is currently not seen in this sample. Nevertheless, some of the CDTGs identified by our analysis may well be associated with dwarf galaxies that have previously merged with the MW, and are now disrupted.

\section{Chemical structure of identified CDTGs}\label{sec:chemical_structure}

\subsection{Statistical Framework}\label{subsec:statistical_framework}

Since the CDTGs we have obtained are expected to contain stars that (within each given CDTG) have been formed in similar primordial environments, the same processes may be responsible for their chemical enrichment. In this section we explore whether the elemental abundances of stars within the found CDTGs are more similar to each other than when the stars are randomly selected from the RPE Final Sample.

We follow the same statistical framework as outlined in \citet{Gudin2021}. First, we use Monte Carlo sampling to select $2.5\times 10^6$ random groups of $N$ stars with a given measured elemental abundance (with $5\leq N\leq 22$) from the sample and measure the biweight scale \citep{Beers1990a}. This allows us to obtain empirical estimates of the cumulative distribution functions (CDFs) for the elemental-abundance dispersions within CDTGs of a given size. A low CDF value for a given elemental-abundance dispersion in a given CDTG indicates increased similarity for this species between the cluster member stars.

For the elemental-abundance dispersions selected at random for CDTGs of a given size, the probability of the number of clusters lying below a given CDF value (from 0 to 1) is described by the binomial distribution. Using three different CDF thresholds ($\alpha\in\{0.25,0.33,0.5\}$) and multiple different abundances ($X\in\{\text{[Fe/H]}, \text{[C/Fe]}_\text{c}, \dots\}$), we obtain overall statistical significances from multinomial distributions obtained by either grouping the cumulative probabilities across all $\alpha$ values, or across all $X$ abundances. The overall statistical significance of the results is obtained by grouping the probabilities across both all $\alpha$ values and all $X$ abundances. We denote these probabilities according to the following classification:
\begin{itemize}
    \item \textit{Individual Elemental-Abundance Dispersion (IEAD) probability}: Individual binomial probability for specific values of $\alpha$ and $X$.
    \item \textit{Full Elemental-Abundance Distribution (FEAD) probability}: Multinomial probability for specific values of $\alpha$, grouped over all abundances $X$.
    \item \textit{Global Element Abundance Dispersion (GEAD) probability}: Multinomial probability for specific abundances $X$, grouped over all values of $\alpha$. This is the overall statistical significance for the particular abundance.
    \item \textit{Overall Element Abundance Dispersion (OEAD) probability}: Multinomial probability grouped over all values of $\alpha$ and all abundances $X$. This is the overall statistical significance of our clustering results.
\end{itemize}

For a more detailed discussion of the above probabilities, and their use, the interested reader is referred to \citet{Gudin2021}. 

\vspace{2cm}

\startlongtable
\begin{deluxetable*}{l  r  c  c  c  c  c  c  c}
\tablecaption{CDTG Abundance Means, Dispersions and Inter-Quartile Ranges (IQRs) \label{tab:cluster_mean}}
\tablehead{\colhead{Cluster} & \colhead{$N$ Stars} & \colhead{[Fe/H]} & \colhead{[C/Fe]$_{c}$} & \colhead{[Mg/Fe]} & \colhead{[Sr/Fe]} & \colhead{[Y/Fe]} & \colhead{[Ba/Fe]} & \colhead{[Eu/Fe]}}
\startdata
CDTG-$1$ & $22$ & $-1.00\pm0.20$ &$+0.52\pm0.06$ &$\dots$ &$+0.64\pm\dots$ &$+0.24\pm0.20$ &$+0.32\pm0.20$ &$+0.51\pm0.10$\\
CDTG-$2$ & $21$ & $-1.33\pm0.75$ &$+0.30\pm0.23$ &$+0.35\pm0.20$ &$+0.15\pm0.35$ &$+0.11\pm0.25$ &$+0.26\pm0.15$ &$+0.57\pm0.17$\\
CDTG-$3$ & $18$ & $-0.87\pm0.07$ &$+0.47\pm0.06$ &$\dots$ &$+0.95\pm\dots$ &$+0.11\pm0.14$ &$+0.13\pm0.19$ &$+0.49\pm0.09$\\
CDTG-$4$ & $18$ & $-1.52\pm0.57$ &$+0.29\pm0.20$ &$+0.24\pm0.21$ &$+0.13\pm0.07$ &$+0.06\pm0.18$ &$+0.26\pm0.27$ &$+0.51\pm0.15$\\
CDTG-$5$ & $17$ & $-1.28\pm0.47$ &$+0.38\pm0.17$ &$+0.32\pm\dots$ &$+0.24\pm0.07$ &$+0.22\pm0.20$ &$+0.37\pm0.26$ &$+0.62\pm0.16$\\
CDTG-$6$ & $16$ & $-0.86\pm0.05$ &$+0.49\pm0.13$ &$+0.45\pm\dots$ &$+0.28\pm0.63$ &$+0.08\pm0.17$ &$+0.11\pm0.11$ &$+0.44\pm0.09$\\
CDTG-$7$ & $16$ & $-1.89\pm0.76$ &$+0.53\pm0.19$ &$+0.44\pm0.13$ &$+0.31\pm0.79$ &$+0.03\pm0.19$ &$+0.14\pm0.28$ &$+0.55\pm0.18$\\
CDTG-$8$ & $14$ & $-1.62\pm0.78$ &$+0.47\pm0.22$ &$+0.37\pm0.03$ &$+0.30\pm0.27$ &$+0.26\pm0.18$ &$+0.33\pm0.41$ &$+0.71\pm0.29$\\
CDTG-$9$ & $12$ & $-0.88\pm0.09$ &$+0.50\pm0.06$ &$+0.44\pm\dots$ &$+1.44\pm0.00$ &$+0.18\pm0.15$ &$+0.17\pm0.21$ &$+0.57\pm0.26$\\
CDTG-$10$ & $12$ & $-0.85\pm0.02$ &$+0.46\pm0.08$ &$\dots$ &$+1.59\pm\dots$ &$+0.14\pm0.16$ &$+0.18\pm0.20$ &$+0.51\pm0.10$\\
CDTG-$11$ & $11$ & $-1.49\pm0.19$ &$+0.54\pm0.10$ &$+0.42\pm0.04$ &$-0.29\pm\dots$ &$+0.24\pm0.09$ &$+0.29\pm0.22$ &$+0.49\pm0.13$\\
CDTG-$12$ & $11$ & $-2.16\pm0.81$ &$+0.53\pm0.35$ &$+0.38\pm0.16$ &$+0.12\pm0.12$ &$-0.02\pm0.31$ &$+0.02\pm0.41$ &$+0.58\pm0.16$\\
CDTG-$13$ & $11$ & $-0.86\pm0.04$ &$+0.49\pm0.07$ &$\dots$ &$+1.13\pm\dots$ &$+0.15\pm0.20$ &$+0.15\pm0.10$ &$+0.52\pm0.10$\\
CDTG-$14$ & $11$ & $-0.91\pm0.14$ &$+0.51\pm0.09$ &$\dots$ &$+1.42\pm\dots$ &$+0.22\pm0.17$ &$+0.25\pm0.20$ &$+0.49\pm0.08$\\
CDTG-$15$ & $10$ & $-1.29\pm0.22$ &$+0.41\pm0.11$ &$+0.31\pm\dots$ &$+1.68\pm\dots$ &$+0.09\pm0.07$ &$+0.39\pm0.23$ &$+0.64\pm0.14$\\
CDTG-$16$ & $10$ & $-0.91\pm0.14$ &$+0.45\pm0.10$ &$\dots$ &$+0.35\pm\dots$ &$+0.15\pm0.13$ &$+0.25\pm0.22$ &$+0.55\pm0.08$\\
CDTG-$17$ & $10$ & $-1.99\pm0.54$ &$+0.20\pm0.37$ &$+0.33\pm0.08$ &$+0.08\pm0.33$ &$-0.18\pm0.17$ &$-0.01\pm0.25$ &$+0.47\pm0.15$\\
CDTG-$18$ & $9$ & $-2.31\pm0.70$ &$+0.39\pm0.22$ &$+0.35\pm0.20$ &$+0.06\pm0.32$ &$+0.15\pm0.12$ &$+0.25\pm0.27$ &$+0.73\pm0.29$\\
CDTG-$19$ & $9$ & $-1.40\pm0.36$ &$+0.37\pm0.26$ &$+0.37\pm\dots$ &$+0.30\pm0.05$ &$-0.02\pm0.36$ &$+0.38\pm0.21$ &$+0.68\pm0.26$\\
CDTG-$20$ & $9$ & $-1.16\pm0.44$ &$+0.37\pm0.09$ &$\dots$ &$+0.08\pm\dots$ &$+0.19\pm0.20$ &$+0.41\pm0.25$ &$+0.64\pm0.13$\\
CDTG-$21$ & $9$ & $-0.98\pm0.31$ &$+0.41\pm0.10$ &$\dots$ &$+0.90\pm\dots$ &$+0.11\pm0.16$ &$+0.25\pm0.26$ &$+0.55\pm0.13$\\
CDTG-$22$ & $8$ & $-2.06\pm0.60$ &$+0.40\pm0.33$ &$+0.54\pm0.17$ &$-0.05\pm0.23$ &$-0.02\pm0.22$ &$+0.11\pm0.24$ &$+0.48\pm0.22$\\
CDTG-$23$ & $8$ & $-0.89\pm0.09$ &$+0.47\pm0.19$ &$+0.48\pm\dots$ &$+0.31\pm0.84$ &$-0.05\pm0.17$ &$+0.08\pm0.32$ &$+0.48\pm0.13$\\
CDTG-$24$ & $8$ & $-0.97\pm0.28$ &$+0.52\pm0.09$ &$+0.35\pm\dots$ &$+0.37\pm\dots$ &$+0.26\pm0.18$ &$+0.26\pm0.28$ &$+0.51\pm0.11$\\
CDTG-$25$ & $8$ & $-2.20\pm0.52$ &$+0.18\pm0.38$ &$+0.30\pm0.40$ &$+0.03\pm0.14$ &$-0.25\pm0.07$ &$+0.06\pm0.30$ &$+0.49\pm0.10$\\
CDTG-$26$ & $7$ & $-0.84\pm0.10$ &$+0.41\pm0.02$ &$\dots$ &$\dots$ &$+0.21\pm0.12$ &$+0.33\pm0.16$ &$+0.48\pm0.05$\\
CDTG-$27$ & $7$ & $-2.03\pm0.65$ &$+0.55\pm0.26$ &$+0.49\pm\dots$ &$+0.15\pm\dots$ &$+0.11\pm0.30$ &$+0.25\pm0.33$ &$+0.38\pm0.05$\\
CDTG-$28$ & $7$ & $-1.31\pm0.52$ &$+0.47\pm0.20$ &$+0.52\pm0.15$ &$+0.15\pm0.41$ &$+0.02\pm0.23$ &$+0.22\pm0.18$ &$+0.58\pm0.17$\\
CDTG-$29$ & $6$ & $-0.90\pm0.09$ &$+0.42\pm0.08$ &$\dots$ &$\dots$ &$+0.06\pm0.17$ &$+0.11\pm0.10$ &$+0.56\pm0.12$\\
CDTG-$30$ & $6$ & $-0.89\pm0.07$ &$+0.43\pm0.04$ &$+0.11\pm\dots$ &$+0.32\pm\dots$ &$-0.02\pm0.17$ &$+0.10\pm0.16$ &$+0.53\pm0.16$\\
CDTG-$31$ & $6$ & $-0.92\pm0.06$ &$+0.47\pm0.04$ &$\dots$ &$\dots$ &$-0.03\pm0.11$ &$+0.13\pm0.08$ &$+0.48\pm0.09$\\
CDTG-$32$ & $6$ & $-0.86\pm0.03$ &$+0.44\pm0.22$ &$+0.32\pm\dots$ &$-0.32\pm0.35$ &$+0.24\pm0.16$ &$+0.11\pm0.17$ &$+0.43\pm0.06$\\
CDTG-$33$ & $6$ & $-1.87\pm0.61$ &$+0.34\pm0.23$ &$+0.31\pm0.13$ &$+0.19\pm0.04$ &$+0.01\pm0.32$ &$+0.18\pm0.41$ &$+0.63\pm0.15$\\
CDTG-$34$ & $6$ & $-1.25\pm0.60$ &$+0.36\pm0.17$ &$+0.40\pm\dots$ &$-0.03\pm\dots$ &$-0.02\pm0.21$ &$+0.49\pm0.04$ &$+0.62\pm0.06$\\
CDTG-$35$ & $5$ & $-0.91\pm0.15$ &$+0.47\pm0.11$ &$\dots$ &$\dots$ &$+0.16\pm0.29$ &$+0.08\pm0.30$ &$+0.56\pm0.08$\\
CDTG-$36$ & $5$ & $-0.91\pm0.09$ &$+0.53\pm0.05$ &$\dots$ &$+0.13\pm\dots$ &$-0.01\pm0.12$ &$+0.08\pm0.12$ &$+0.53\pm0.13$\\
\cline{0-8}
Biweight (CDTG mean): & & $-1.11\pm0.08$ & $+0.45\pm0.01$ & $+0.38\pm0.02$ & $+0.20\pm0.06$ & $+0.10\pm0.02$ & $+0.20\pm0.02$ & $+0.53\pm0.01$\\
Biweight (CDTG std): & & $+0.31\pm0.05$ & $+0.14\pm0.02$ & $+0.14\pm0.02$ & $+0.26\pm0.06$ & $+0.17\pm0.01$ & $+0.22\pm0.01$ & $+0.13\pm0.01$\\
IQR (CDTG mean): & & $\boldsymbol{0.65}$ & $\boldsymbol{0.11}$ & $\boldsymbol{0.12}$ & $\boldsymbol{0.33}$ & $\boldsymbol{0.20}$ & $\boldsymbol{0.16}$ & $\boldsymbol{0.09}$\\
IQR (CDTG std): & & $0.49$ & $0.14$ & $0.08$ & $0.28$ & $0.06$ & $0.11$ & $0.07$\\
\cline{0-8}
Biweight (Final): & & $-1.27\pm0.68$ & $+0.45\pm0.19$ & $+0.37\pm0.20$ & $+0.14\pm0.38$ & $+0.10\pm0.21$ & $+0.23\pm0.28$ & $+0.55\pm0.16$\\
IQR (Final): & & $\boldsymbol{1.08}$ & $\boldsymbol{0.22}$ & $\boldsymbol{0.24}$ & $\boldsymbol{0.40}$ & $\boldsymbol{0.28}$ & $\boldsymbol{0.36}$ & $\boldsymbol{0.21}$\\
\enddata
\tablecomments{The first section of the table lists the biweight location and scale of the abundances for each of the CDTGs. The second section of the table lists the mean and the standard error of the mean (using biweight estimates) for both the location and scale of the abundances of the CDTGs, along with the IQR of the abundances of the CDTGs. The third section of the table lists the biweight location and scale of the Final Sample for each of the abundances, along with the IQR for each of the abundances in the Final Sample. The IQR lines are bolded to draw attention to the CDTG 
elemental-abundance mean IQRs to compare to the Final Sample abundance IQRs. }\end{deluxetable*}

\begin{deluxetable*}{cccccc}[!]
\tabletypesize{\footnotesize}
\tablecaption{CDTG Elemental-Abundance Statistics: Final Sample\label{tab:binomial_probability_table}}
\tablehead{\colhead{Abundance} & \colhead{$\#$ CDTGs} & \colhead{$  N < 0.5$, $0.33$, $0.25$} & \colhead{IEAD Probabilities} & \colhead{GEAD Probabilities} & \colhead{OEAD Probability}}
\startdata
\text{[Fe/H]} & 36 & $24$, $20$, $18$& $\phantom{1}3.3\%$, $\phantom{1}0.4\%$, $\phantom{1}0.1\%$,  & $\phantom{1}0.1\%$ & \hfil\multirow{7}{*}{$\ll0.01\%$}\hfill\\
\text{[C/Fe]$_{c}$} & 36 & $21$, $16$, $15$& $20.3\%$, $10.1\%$, $\phantom{1}2.1\%$,  & $\phantom{1}1.6\%$ & \\
\text{[Mg/Fe]} & 12 & $\phantom{1}5$, $\phantom{1}3$, $\phantom{1}2$& $80.6\%$, $81.2\%$, $84.2\%$,  & $69.1\%$ & \\
\text{[Sr/Fe]} & 17 & $\phantom{1}9$, $\phantom{1}6$, $\phantom{1}6$& $50.0\%$, $51.1\%$, $23.5\%$,  & $19.7\%$ & \\
\text{[Y/Fe]} & 36 & $23$, $13$, $10$& $\phantom{1}6.6\%$, $40.6\%$, $41.2\%$,  & $\phantom{1}5.3\%$ & \\
\text{[Ba/Fe]} & 36 & $21$, $17$, $16$& $20.3\%$, $\phantom{1}5.3\%$, $\phantom{1}0.9\%$,  & $\phantom{1}0.7\%$ & \\
\text{[Eu/Fe]} & 36 & $22$, $16$, $14$& $12.1\%$, $10.1\%$, $\phantom{1}4.6\%$,  & $\phantom{1}2.3\%$ & \\
\hline
\multicolumn{2}{c}{FEAD Probabilities} & & $0.09\%$, $0.01\%$, $\ll0.01\%$ & &
\enddata
\tablecomments{The Individual Elemental-Abundance Dispersion (IEAD) probabilities represent the binomial probabilities for each element for the 0.5, 0.33, and 0.25, respectively. The Full Elemental-Abundance Dispersion (FEAD) probabilities represent the probabilities (across {\it all} elements) for the 0.5, 0.33, and 0.25, respectively. The Global Elemental-Abundance Dispersion (GEAD) probabilities represent the probabilities for the triplet of CDF levels for each element. The Overall Elemental-Abundance Dispersion (OEAD) probability represents the probability (across {\it all} elements) resulting from random draws from the full CDF.  See text for details.}
\end{deluxetable*}

\subsection{Important Caveats}

Note that there are several caveats to the interpretation of this scheme that should be mentioned. Most importantly, a meaningful understanding of the derived probabilities (described below) depends on the contrast of the ``typical" CDTG elemental-abundance dispersion to the abundance dispersion of the parent population to which it is compared, from which random draws are made. If the typical CDTG elemental-abundance dispersion is roughly commensurate with the dispersion of the parent sample, then by definition the dispersions will always be consistent with what are expected from the random draws, and one cannot expect the CDTG elemental-abundance dispersions to be significantly smaller. This requires that, in particular for a given element, prior to assessing the significance of its dispersion for a given set of CDTGs, we should compare its value to the expected value from the appropriate parent sample. We carry out this exercise, as described below, by inspection of the Inter-Quartile Range (IQR) for each element for a given set of CDTGs compared to the IQR of CDTGs drawn at random from the parent sample. As a rule of thumb, we demand that the IQR of the mean for each element of a set of CDTGs is on the order of one-half of the IQR of the parent-population CDTGs. Otherwise, there is insufficient ``dynamical range" for the statistical inferences to be made with confidence, at least for individual elements.  

Furthermore, as is perhaps obvious, the statistical power of our comparisons increase with the numbers of CDTGs in a given parent population. Thus, when we consider dynamical clusters formed exclusively from $r$-II stars, as discussed below, it becomes more difficult to place confidence in the statistical inferences.  For that reason, in the case of the clustering of $r$-II stars, we have relaxed the criterion for the minimum number of stars per cluster to $3$, rather than $5$, more than doubling the numbers of identified $r$-II CDTGs. 

%Nevertheless, we believe the methodology is sound, and its power will increase as sample sizes of minority populations (such as the $r$-II stars) grow in the future. 

It should also be kept in mind that the observed CDTG elemental-abundance dispersions depend on a number of different parameters, including not only on the total mass of a given parent dwarf galaxy, but on its available gas mass for conversion into stars, the history of star formation in that environment, and the nature of the progenitor population(s) involved in the production of a given element. These are complex and interacting sets of conditions, and certainly are best considered in the context of simulations (such as \citealt{Hirai2022}). Consequently, the expected result for a given element in a given set of CDTGs is not always clear. However, we have designed our statistical tests to consider a broad set of questions of interest, the most pertinent of which for the current application are the FEAD and OEAD probabilities, which we employ for making our primary inferences. 

%As described above, this probability describes the statistical likelihood of the overall distributions of dispersions for all elements over the CDTG sample being considered.

\begin{deluxetable*}{cccccc}[!]
\tabletypesize{\footnotesize}
\tablecaption{CDTG Elemental-Abundance Statistics: $r$-I Sample\label{tab:binomial_probability_rI_all_table}}
\tablehead{\colhead{Abundance} & \colhead{$\#$ CDTGs} & \colhead{$  N < 0.5$, $0.33$, $0.25$} & \colhead{IEAD Probabilities} & \colhead{GEAD Probabilities} & \colhead{OEAD Probability}}
\startdata
\text{[Fe/H]} & 28 & $16$, $14$, $13$& $28.6\%$, $\phantom{1}4.6\%$, $\phantom{1}1.1\%$,  & $\phantom{1}1.0\%$ & \hfil\multirow{7}{*}{$\ll0.01\%$}\hfill\\
\text{[C/Fe]$_{c}$} & 28 & $15$, $14$, $14$& $42.5\%$, $\phantom{1}4.6\%$, $\phantom{1}0.4\%$,  & $\phantom{1}0.4\%$ & \\
\text{[Mg/Fe]} & \phantom{1}8 & $\phantom{1}6$, $\phantom{1}5$, $\phantom{1}4$& $14.5\%$, $\phantom{1}8.5\%$, $11.4\%$,  & $\phantom{1}4.1\%$ & \\
\text{[Sr/Fe]} & 12 & $\phantom{1}5$, $\phantom{1}5$, $\phantom{1}5$& $80.6\%$, $35.9\%$, $15.8\%$,  & $15.8\%$ & \\
\text{[Y/Fe]} & 27 & $19$, $12$, $\phantom{1}9$& $\phantom{1}2.6\%$, $14.5\%$, $21.4\%$,  & $\phantom{1}1.6\%$ & \\
\text{[Ba/Fe]} & 28 & $19$, $14$, $12$& $\phantom{1}4.4\%$, $\phantom{1}4.6\%$, $\phantom{1}2.9\%$,  & $\phantom{1}0.8\%$ & \\
\text{[Eu/Fe]} & 28 & $15$, $12$, $\phantom{1}9$& $42.5\%$, $18.1\%$, $25.0\%$,  & $12.6\%$ & \\
\enddata
\tablecomments{The Individual Elemental-Abundance Dispersion (IEAD) probabilities represent the binomial probabilities for each element for the 0.5, 0.33, and 0.25, respectively. The Full Elemental-Abundance Dispersion (FEAD) probabilities represent the probabilities (across {\it all} elements) for the 0.5, 0.33, and 0.25, respectively. The Global Elemental-Abundance Dispersion (GEAD) probabilities represent the probabilities for the triplet of CDF levels for each element. The Overall Elemental-Abundance Dispersion (OEAD) probability represents the probability (across {\it all} elements) resulting from random draws from the full CDF.  See text for details.}
\end{deluxetable*}

\subsection{Results}

\subsubsection{Full Sample of RPE CDTGs}

Table~\ref{tab:cluster_mean} lists the means and dispersions of the elemental abundances explored in this study for each of the CDTGs identified in this work. The second part of the table lists the global CDTG properties, with the mean and standard error of the mean (using biweight location and scale) of both the CDTG means and dispersions being listed. The second part of the table also includes the IQR of the CDTG means and dispersions. The third part of the table lists the biweight location and scale of the elemental abundances in the Final Sample, along with the IQR of the elemental abundances in the Final Sample. As can be seen by comparing the two IQRs for each the CDTG results and the Final Sample, the IQRs for the CDTG results for 4 of the 7 elements considered are at least twice smaller, the exceptions being for [Fe/H] (which only slightly misses our rule of thumb), [Sr/Fe], and [Y/Fe]. The elements that meet the criteria, [C/Fe]$_{c}$, [Mg/Fe], [Ba/Fe], and [Eu/Fe], provide stronger constraints compared to the ones that have large IQR ranges.

Table~\ref{tab:binomial_probability_table} lists the numbers of CDTGs with available estimates of the listed abundance ratios, and the numbers of CDTGs falling below the $0.50$, $0.33$, and $0.25$ levels of the CDFs, along with our calculated values for the various probabilities.  The full and overall probabilities (captured by the FEAD and OEAD probability values) are very low, implying that the measured abundance spreads are highly statistically significant within our CDTGs across the entire sample; this is similar to the results found in \citet{Gudin2021} (see their Table 7). However, some of the individual abundance spreads (the GEAD probabilities) are not statistically significant, namely the [Mg/Fe] and [Sr/Fe] spreads. Keep in mind, as noted above, that the lack of contrast in the mean IQRs of [Fe/H], [Sr/Fe], and [Y/Fe] for our CDTGs with their corresponding mean IQRs for the full sample may impact interpretation of their probabilities. 

By comparison, in \citet{Gudin2021}, the [Fe/H], [C/Fe]$_\text{c}$, [Sr/Fe], and [Ba/Fe] spreads were statistically significant, as inferred from their GEAD probabilities; the same is recovered in this study, with the exception of [Sr/Fe].  It should also be recalled that the sample of RPE stars considered by Gudin et al. is dominated by stars with lower [Fe/H] than our present sample, since the generally more metal-rich stars from GALAH were not available at that time.

%In addition to performing the statistical analysis on the full set of abundances ([Fe/H], [C/Fe]$_\text{c}$, [Mg/Fe], [Sr/Fe], [Y/Fe], [Ba/Fe] and [Eu/Fe]), we group light $r$-process elements ([Sr/Fe], [Y/Fe]) and heavy $r$-process elements ([Ba/Fe],[Eu/Fe]) into combined abundances (denoted as [lr/Fe] and [hr/Fe], respectively), by taking the averages of the stellar abundances for stars in which both of the respective elements are detected. Table~\ref{tab:binomial_probability_combined_table} shows how this procedure affects the entire sample. It is clear that, for the full sample, the FEAD probabilities generally become statistically insignificant, while the OEAD probability remains significant.

Below we explore whether $r$-I and $r$-II stars exhibit different elemental-abundance patterns. For this purpose, we repeat the clustering procedure described in Section \ref{sec:ClusteringProcedure}, separately for subsets of $r$-I and $r$-II stars. The result of this procedure is $28$ ($r$-I) CDTGs ranging in size from $5$ to $18$ members and $20$ ($r$-II) CDTGs ranging in size from $3$ to $23$ members, respectively. We then perform the same statistical analysis on each as described above. 

\begin{deluxetable*}{cccccc}[!]
\tabletypesize{\footnotesize}
\tablecaption{CDTG Elemental-Abundance Statistics: $r$-II Sample\label{tab:binomial_probability_rII_all_3_table}}
\tablehead{\colhead{Abundance} & \colhead{$\#$ CDTGs} & \colhead{$  N < 0.5$, $0.33$, $0.25$} & \colhead{IEAD Probabilities} & \colhead{GEAD Probabilities} & \colhead{OEAD Probability}}
\startdata
\text{[Fe/H]} & 20 & $16$, $14$, $11$& $\phantom{1}0.6\%$, $\phantom{1}0.1\%$, $\phantom{1}0.4\%$,  & $\phantom{1}0.0\%$ & \hfil\multirow{7}{*}{$\ll0.01\%$}\hfill\\
\text{[C/Fe]$_{c}$} & 19 & $13$, $11$, $\phantom{1}9$& $\phantom{1}8.4\%$, $\phantom{1}2.2\%$, $\phantom{1}2.9\%$,  & $\phantom{1}0.9\%$ & \\
\text{[Mg/Fe]} & \phantom{1}5 & $\phantom{1}2$, $\phantom{1}1$, $\phantom{1}1$& $81.2\%$, $86.5\%$, $76.3\%$,  & $68.5\%$ & \\
\text{[Sr/Fe]} & 10 & $\phantom{1}5$, $\phantom{1}3$, $\phantom{1}3$& $62.3\%$, $69.3\%$, $47.4\%$,  & $39.6\%$ & \\
\text{[Y/Fe]} & 17 & $11$, $\phantom{1}7$, $\phantom{1}6$& $16.6\%$, $31.5\%$, $23.5\%$,  & $\phantom{1}9.0\%$ & \\
\text{[Ba/Fe]} & 20 & $12$, $10$, $\phantom{1}5$& $25.2\%$, $\phantom{1}8.7\%$, $58.5\%$,  & $\phantom{1}7.2\%$ & \\
\text{[Eu/Fe]} & 20 & $12$, $\phantom{1}9$, $\phantom{1}8$& $25.2\%$, $18.2\%$, $10.2\%$,  & $\phantom{1}6.5\%$ & \\
\enddata
\tablecomments{The Individual Elemental-Abundance Dispersion (IEAD) probabilities represent the binomial probabilities for each element for the 0.5, 0.33, and 0.25, respectively. The Full Elemental-Abundance Dispersion (FEAD) probabilities represent the probabilities (across {\it all} elements) for the 0.5, 0.33, and 0.25, respectively. The Global Elemental-Abundance Dispersion (GEAD) probabilities represent the probabilities for the triplet of CDF levels for each element. The Overall Elemental-Abundance Dispersion (OEAD) probability represents the probability (across {\it all} elements) resulting from random draws from the full CDF.  See text for details.}
\end{deluxetable*}

% The expectation, in the case that $r$-I and $r$-II stars have different chemical-enrichment histories, is that the light $r$-process element abundances will be significantly reduced in the $r$-I CDTGs, and the heavy $r$-process element abundances will be reduced in the $r$-II CDTGs \citep{Hirai2019}. 

\subsubsection{The r-I Sample}\label{subsec:rI_sample}
The binomial statistics for the $28$ $r$-I CDTGs are listed in Table~\ref{tab:binomial_probability_rI_all_table}.  We observe a lack of statistical significance for the reduction of [Sr/Fe] and [Eu/Fe] abundance spreads, but other abundances (including [Mg/Fe], which exhibited a statistically insignificant spread reduction in the case of the full sample) exhibit statistically significant reductions in their spreads (all elements except for [Sr/Fe] and [Eu/Fe] have GEAD probabilities less than $10\%$).  

From inspection of Table \ref{tab:cluster_mean_rI} in the Appendix, only 3 of the 7 elements considered ([Y/Fe], [Ba/Fe], and [Eu/Fe]) have contrasts in their mean CDTG IQRs relative to the mean IQRs of the full sample of $r$-I stars that pass our factor of two rule of thumb (although [Fe/H] and [C/Fe]$_\text{c}$ only narrowly miss), so interpretation of the GEAD probabilities for several of these elements should be regarded with caution.

The FEAD probabilities for the $r$-I CDTGs are low, and are statistically significant for all three of the $v = 0.5$, $v = 0.33$ and $v = 0.25$ levels. The OEAD probability is highly statistically significant.

%Table \ref{tab:binomial_probability_rI_combined_table} shows the same statistics with the light and heavy $r$-process elements grouped into [lr/Fe] and [hr/Fe], respectively. The overall reduction in the abundance spreads is not statistically significant (the OEAD probability is well above $5\%$), and it is not for any of the individual abundance groups (all GEAD probabilities are $\gtrsim 15\%$).

\subsubsection{The r-II Sample}\label{subsec:rII_sample}
%One caveat of the following analysis is that for the $r$-II sub-sample, we only have $7$ CDTGs. The stability of the results generally drops with the decrease of the number of clusters, as the outlier abundance values in individual clusters begin to affect the measured CDF values. As the number of known $r$-II stars with accurately measured $r$-process element abundances increases in the future, revisions of this analysis are expected to be marked by an improvement in stability. To perform this task in the future with less variance, the number of $r$-II stars will have to reach $\sim 500$, which was the approximate total number of $r$-I and $r$-II stars used by \citet{Gudin2021} when employing these statistical techniques.

If we adopt the minimum number of stars per $r$-II CDTG for the clustering exercise as we have for the full sample ($5$) and the $r$-I sample ($5$), we are left with only a total of $7$ $r$-II CDTGs for the statistical analysis.  We judge this to be too small, as experience suggests that a minimum of $10$ CDTGs provide the most stable results.
Thus, we chose to reduce the minimum number of stars in order to form an $r$-II CDTG from $5$ to $3$. This increases the number of $r$-II CDTGs from $7$ to $20$, comparable to that in the full and the $r$-I samples ($36$ and $28$ respectively).

Table \ref{tab:binomial_probability_rII_all_3_table} shows the binomial statistics for our $20$ $r$-II CDTGs. From inspection, the GEAD probabilities for the elements considered are statistically significant in the same manner as the Final Sample, with [Mg/Fe] and [Sr/Fe] both lacking stastical significance. 

From inspection of Table \ref{tab:cluster_mean_rII} in the Appendix, only $2$ of the $7$ elements considered ([Y/Fe] and [Ba/Fe])  have mean IQRs for the $r$-II CDTGs that are at least twice smaller than the mean IQR of the final sample, although [C/Fe]$_\text{c}$ only narrowly misses our rule of thumb.  Thus we can assume that, for many of the individual elements, there is not a useful dynamical range for confident interpretation for most of the GEADs.

The FEAD probabilities for the $r$-II CDTGs are statistically significant for all three $v = 0.50$, $v = 0.33$, and $v = 0.25$ levels. The OEAD probability is highly statistically significant.

\section{Summary}\label{sec:Discussion_2}

We have assembled an RPE Initial Sample of $1776$ stars ($1393$ $r$-I stars, $381$ $r$-II stars, and $2$ $r$-III stars) from both a literature search and GALAH DR3 \citep{Buder2021} survey data, with stars that met the $r$-process-enhancement requirements listed in Table~\ref{tab:MPsignatures}. A total of $105$ of these stars are identified as CEMP-$r$ stars; these are listed in Table~\ref{tab:cemp} in the Appendix. These stars are of interest due to their enhanced carbon abundance and association to the morphological groups described in \citet{Yoon2016}. Based on their classification scheme, there are $58$ Group I CEMP-$r$ stars, $41$ Group II CEMP-$r$ stars, and $2$ Group III CEMP-$r$ stars, with a number of stars that have ambiguous classifications. This list provides a useful reference for high-resolution follow-up targets, some of which has already begun (e.g., \citealt{Rasmussen2020} and \citealt{Zepeda2022}).

The RPE Final Sample of $1720$ stars ($1346$ $r$-I stars, $372$ $r$-II stars, and $2$ $r$-III stars) had radial velocity and astrometric information from which orbits were constructed, in order to identify Chemo-Dynamically Tagged Groups (CDTGs) in orbital energy and cylindrical action space with the \HDBSCAN ~algorithm. We chose \HDBSCAN\ as the clustering algorithm due to precedence within the literature \citep{Koppelman2019a,Gudin2021,Limberg2021a,Shank2022a}, and its ability to extract clusters of stars over the energy and action space. 

%Other clustering algorithms have been considered in the past, such as agglomerative clustering, affinity propogation, K-means, and mean-shift clustering \citep{Roederer2018a}, along with friends-of-friends \citep{Gudin2021}. 

We recover $36$ CDTGs that include between $5$ and $22$ members, with $17$ CDTGs containing at least $10$ member stars. These CDTGs were associated with MW substructures, resulting in the re-identification of the Gaia-Sausage-Enceladus, the Splashed Disk, the Metal-Weak Thick Disk, the Helmi Stream, LMS-1 (Wukong), and Thamnos. A total of $7$ CDTGs were associated with globular clusters, while no surviving dwarf galaxies were determined to be associated with the identified CDTGs. Previously identified groups were found to be associated with the CDTGs as well, with past work mostly confirming our substructure identification, and showing some limitations in the procedure, which we discussed. Each of these associations allow insights into the dynamical and chemical properties of the parent substructures. 

The implications of past group and stellar associations were explored, with emphasis placed on the structure associations. Stellar associations to stars with abnormal MW abundances were addressed as being good candidates for high-resolution follow-up spectroscopy targets, due to the statistical likelihood of the other members being chemically peculiar as well, mostly focused on RPE and CEMP stars. 

We have considered the statistical significance of the elemental-abundance dispersions across the identified RPE CDTGs for a set of seven abundance ratios ([Fe/H]], [C/Fe]$_\text{c}$, [Mg/Fe], [Sr/Fe], [Y/Fe], [Ba/Fe], and [Eu/Fe]). CDTGs are statistically examined, in order to assess the similarity (or not) of the chemical-evolution histories in their presumed dwarf-galaxy birth environments, following the approach developed by \citet{Gudin2021}.  We point out that, for a number of elements considered, the mean IQRs of the dispersions in the CDTGs do not provide sufficient dynamical range compared to the mean IQR of the parent samples with which they are compared (we suggest they must be a factor of two smaller) in order to enable meaningful interpretations of the Global Elemental-Abundance (GEAD) probabilities. 

However, the probabilities that consider the distributions of {\it all} of the elements across the CDFs for the set of CDTGs (the Full Elemental-Abundance Distribution, FEAD probabilities, and the Overall Elemental-Abundance Distribution, OEAD probablities), strongly support the assertion that the stars associated with individual RPE CDTGs indeed share similar chemical-enrichment histories, as previously claimed by \citet{Gudin2021}.  We have also divided the full sample into RPE stars classified as moderately $r$-process enhanced ($r$-I) and highly $r$-process enhanced ($r$-II), and find similar results.

The methods presented here will be used in future samples that contain many more RPE stars, especially with planned data releases from the RPA increasing their total number, and other ongoing or planned surveys allowing for a systematic search for RPE stars.  The next steps in advancing our understanding of the birth environments and the nature of the astrophysical site(s) of the $r$-process require detailed comparisons with modern high-resolution (spatial and temporal) simulations of the formation of Milky Way-like galaxies, along the lines explored by \citet{Hirai2022}.

%The methods presented here will be used on larger samples of field stars that we are in the process of assembling -- the HK/HES/HKII surveys \citep{Beers1985,Beers1992,Christlieb2008,Rhee2001} (a subset of which were analyzed by \citealt{Limberg2021a}), as well as SDSS/LAMOST and APOGEE. These data sets will also be supplemented with photometric estimates of effective temperature and metallicity from \citet{Huang2022}, which allows stars from the HK/HES/HKII surveys with no previous spectroscopic follow-up to be explored, expanding the data set used by \citet{Limberg2021a}. 

%\input{Tables/binomial_probability_eu_cut_full}
%\input{Tables/binomial_probability_high_metallicity_full}
%\input{Tables/binomial_probability_low_metallicity_full}
%\input{Tables/binomial_probability_eu_cut_rI}
%\input{Tables/binomial_probability_high_metallicity_rI}
%\input{Tables/binomial_probability_low_metallicity_rI}
%\input{Tables/binomial_probability_eu_cut_rII}
%\input{Tables/binomial_probability_high_metallicity_rII}
%\input{Tables/binomial_probability_low_metallicity_rII}

\vspace{2.0cm}

\begin{acknowledgements}

The authors are thankful to an anonymous referee, who provided comments and suggestions that substantially improved the presentation in this paper. We are grateful for data provided by Rohan Naidu that is included in this work. D.S., T.C.B., and I.U.R. acknowledge partial support for this work from grant PHY 14-30152; Physics Frontier Center/JINA Center for the Evolution of the Elements (JINA-CEE), awarded by the US National Science Foundation. The work of V.M.P. is supported by NOIRLab, which is managed by the Association of Universities for Research in Astronomy (AURA) under a cooperative agreement with the US National Science Foundation. I.U.R.\ acknowledges support from the NASA Astrophysics Data Analysis Program, grant 80NSSC21K0627, and the US National Science Foundation, through grant AST~1815403/1815767.

\end{acknowledgements}

\section{Appendix}

Here we present the table for the RPE Initial Sample (Table~\ref{tab:initial_data_descript}) and the RPE Final Sample (Table~\ref{tab:final_data_descript}). 
In the print edition, only the table descriptions are provided; the full tables are available only in electronic form.

We also present Table~\ref{tab:cemp}, which describes the identified CEMP-$r$ stars and their associated morphological groups, according to the regions defined by \citet{Yoon2016}. Tables~\ref{tab:cluster_results_rI_stub} and \ref{tab:cluster_orbit_rI} show the CDTGs identified by \HDBSCAN~ and the CDTG dynamical parameters as determined by \AGAMA~ for the $r$-I Sample, respectively. Table~\ref{tab:cluster_mean_rI} lists the means and dispersions of the CDTGs identified in the $r$-I Sample, along with the statistics on all the CDTGs and the $r$-I Sample. Tables~\ref{tab:cluster_results_rII_stub} and \ref{tab:cluster_orbit_rII} show the CDTGs identified by \HDBSCAN~ and the CDTG dynamical parameters as determined by \AGAMA~ for the $r$-II Sample, respectively. Table~\ref{tab:cluster_mean_rII} lists the means and dispersions of the CDTGs identified in the $r$-II Sample, along with the statistics on all the CDTGs and the $r$-II Sample.

\startlongtable
\begin{deluxetable*}{c  l  l  l}
\tablecaption{Description of the Initial Sample from the RPE Sample \label{tab:initial_data_descript}}
\tablehead{\colhead{Column} & \colhead{Field} & \colhead{Unit} & \colhead{Description}}
\startdata
1 & Name & $-$ & The name of the star as given by the Reference\\
2 & Source ID & $-$ & The Gaia EDR3 Source ID of the star\\
3 & RA & (J2000) & The Right Ascension of the star given in hours:minutes:seconds\\
4 & DEC & (J2000) & The Declination of the star given in degrees:minutes:seconds\\
5 & $V_{\rm mag}$ & $-$ & The $V$ magnitude of the star as given by the $V_{\rm mag}$ Reference\\
6 & $G_{\rm mag}$ & $-$ & The Gaia $G$ mean magnitude of the star as given by the Gaia Source ID\\
7 & $G_{\rm BP} - G_{\rm RP}$ & $-$ & The Gaia BP $-$ RP color mean magnitude of the star as given by the Gaia Source ID\\
8 & $V_{\rm mag}$ (Gaia) & $-$ & The $V$ magnitude of the star as determined by the transformations from G mag to V mag using\\
 &  &  & $V$ $=$ $G$ $+$ $0.02704$ $-$ $0.01424*(\rm{BP}-\rm{RP})$ $+$ $0.2156*(\rm{BP}-\rm{RP})^2$ $-$\\
 &  &  & $0.01426(\rm{BP}-\rm{RP})^3$ given by \citet{Riello2021}\\
9 & RV & (km s$^{-1}$) & The radial velocity as given by RV Reference\\
10 & Error & (km s$^{-1}$) & The radial velocity error as given by RV Reference\\
11 & RV$_{\rm Gaia}$ & (km s$^{-1}$) & The radial velocity as given by the Gaia Source ID\\
12 & Error & (km s$^{-1}$) & The radial velocity error as given by the Gaia Source ID\\
13 & Parallax & (mas) & The parallax as given by the Gaia Source ID\\
14 & Error & (mas) & The parallax error as given by the Gaia Source ID\\
15 & Distance & (kpc) & The inverse parallax distance (1/Parallax)\\
16 & Error & (kpc) & The inverse parallax distance error (Parallax$_{error}$/(Parallax$^2$))\\
17 & Distance$_{\rm Corrected}$ & (kpc) & The corrected inverse parallax distance (1/(Parallax + 0.026)) based on \citet{Huang2021b}\\
18 & Error & (kpc) & The corrected inverse parallax distance error (Parallax$_{error}$/((Parallax + 0.026)$^2$))\\
 &  &  & based on \citet{Huang2021b}\\
19 & Relative Error & $-$ & The relative error of the corrected distance as given by Gaia\\
20 & Distance BJ21 & (kpc) & The 50 percentile distance as given by \citet{Bailer-Jones2021} based on the Gaia Source ID\\
21 & Error & (kpc) & The 50 percentile error as estimated by the 84 percentile distance and the 16 percentile\\
 &  &  & distance as given by \citet{Bailer-Jones2021} based on the Gaia Source ID ((dist84-dist16)/2)\\
22 & Relative Error & $-$ & The relative error of the 50 percentile distance as given by \citet{Bailer-Jones2021} based on\\
 &  &  & the Gaia Source ID\\
23 & Distance StarHorse & (kpc) & The 50 percentile distance as given by \citet{Anders2022} based on the Gaia Source ID\\
24 & Error & (kpc) & The 50 percentile error as estimated by the 84 percentile distance and the 16 percentile\\
 &  &  & distance as given by \citet{Anders2022} based on the Gaia Source ID ((dist84-dist16)/2)\\
25 & Relative Error & $-$ & The relative error of the 50 percentile distance as given by \citet{Anders2022} based on the\\
 &  &  & Gaia Source ID\\
26 & PM$_{\rm RA}$ & (mas yr$^{-1}$) & The proper motion in the Right Ascension as given by the Gaia Source ID\\
27 & Error & (mas yr$^{-1}$) & The proper motion error in the Right Ascension as given by the Gaia Source ID\\
28 & PM$_{\rm DEC}$ & (mas yr$^{-1}$) & The proper motion in the Declination as given by the Gaia Source ID\\
29 & Error & (mas yr$^{-1}$) & The proper motion error in the Declination as given by the Gaia Source ID\\
30 & Correlation Coefficient & $-$ & The correlation coefficient between the proper motion in Right Ascension and the proper motion\\
 &  &  & in Declination as given by the Gaia Source ID\\
31 & T$_{\rm eff}$ & (K) & The effective temperature of the star as given by the Reference\\
32 & log \textit{g} & (cgs) & The surface gravity of the star as given by the Reference\\
33 & [Fe/H] & $-$ & The metallicity of the star as given by (log $\epsilon$(Fe) - log $\epsilon$(Fe)$_\odot$)\\
 &  &  & (Solar value of 7.50 taken from \citet{Asplund2009})\\
34 & log $\epsilon$(Fe) & $-$ & The logarithmic iron abundance of the star as given by the Reference\\
35 & [C/Fe] & $-$ & The carbon abundance ratio of the star as given by (log $\epsilon$(C) - log\\
 &  &  & $\epsilon$(C)$_\odot$ - [Fe/H]) (Solar value of 8.43 taken from \citet{Asplund2009})\\
36 & log $\epsilon$(C) & $-$ & The logarithmic carbon abundance of the star as given by the Reference\\
37 & [C/Fe]$_{c}$ & $-$ & The carbon abundance ratio corrected for evolutionary effects from \citet{Placco2014b}\\
38 & AC$_{c}$ & $-$ & The absolute carbon corrected for evolutionary effects from \citet{Placco2014b} ([C/Fe]$_{c}$\\
 &  &  & + [Fe/H] + log $\epsilon$(C)$_\odot$) (Solar value of 8.43 taken from \citet{Asplund2009})\\
39 & CARDET & $-$ & Flag with ``D" if the carbon abundance ratio ([C/Fe]) is detected by Reference and ``U" if an\\
 &  &  & upper limit by Reference and ``L" if a lower limit by Reference and ``X" if none is detected\\
 &  &  & by Reference\\
40 & CEMP & $-$ & Flag as ``C" for Carbon-Enhanced Metal-Poor (CEMP) if [C/Fe]$_{c}$ $> +0.7$ and ``I" for\\
 &  &  & CEMP-intermediate if $+0.5 <$ [C/Fe]$_{c}$ $\leq +0.7$ and ``N" for Carbon-Normal if\\
 &  &  & [C/Fe]$_{c}$ $\leq +0.5$ and ``X" if there is no [C/Fe]$_{c}$ information\\
41 & [Mg/Fe] & $-$ & The magnesium abundance ratio of the star as given by (log $\epsilon$(Mg) - log\\
 &  &  & $\epsilon$(Mg)$_\odot$ - [Fe/H]) (Solar value of 7.60 taken from \citet{Asplund2009})\\
42 & log $\epsilon$(Mg) & $-$ & The logarithmic magnesium abundance of the star as given by the Reference\\
43 & MAGDET & $-$ & Flag with ``D" if the magnesium abundance ratio ([Mg/Fe]) is detected by Reference and ``U" if\\
 &  &  & an upper limit by Reference and ``L" if a lower limit by Reference and ``X" if none is\\
 &  &  & detected by Reference\\
44 & [Sr/Fe] & $-$ & The strontium abundance ratio of the star as given by (log $\epsilon$(Sr) - log\\
 &  &  & $\epsilon$(Sr)$_\odot$ - [Fe/H]) (Solar value of 2.87 taken from \citet{Asplund2009})\\
45 & log $\epsilon$(Sr) & $-$ & The logarithmic strontium abundance of the star as given by the Reference\\
46 & STRDET & $-$ & Flag with ``D" if the strontium abundance ratio ([Sr/Fe]) is detected by Reference and ``U" if\\
 &  &  & an upper limit by Reference and ``L" if a lower limit by Reference and ``X" if none is\\
 &  &  & detected by Reference\\
47 & [Y/Fe] & $-$ & The yttrium abundance ratio of the star as given by (log $\epsilon$(Y) - log\\
 &  &  & $\epsilon$(Y)$_\odot$ - [Fe/H]) (Solar value of 2.21 taken from \citet{Asplund2009})\\
48 & log $\epsilon$(Y) & $-$ & The logarithmic yttrium abundance of the star as given by the Reference\\
49 & YTTDET & $-$ & Flag with ``D" if the yttrium abundance ratio ([Y/Fe]) is detected by Reference and ``U" if an\\
 &  &  & upper limit by Reference and ``L" if a lower limit by Reference and ``X" if none is detected\\
 &  &  & by Reference\\
50 & [Ba/Fe] & $-$ & The barium abundance ratio of the star as given by (log $\epsilon$(Ba) - log\\
 &  &  & $\epsilon$(Ba)$_\odot$ - [Fe/H]) (Solar value of 2.18 taken from \citet{Asplund2009})\\
51 & log $\epsilon$(Ba) & $-$ & The logarithmic barium abundance of the star as given by the Reference\\
52 & BARDET & $-$ & Flag with ``D" if the barium abundance ratio ([Ba/Fe]) is detected by Reference and ``U" if an\\
 &  &  & upper limit by Reference and ``L" if a lower limit by Reference and ``X" if none is detected\\
 &  &  & by Reference\\
53 & [Eu/Fe] & $-$ & The europium abundance ratio of the star as given by (log $\epsilon$(Eu) - log\\
 &  &  & $\epsilon$(Eu)$_\odot$ - [Fe/H]) (Solar value of 0.52 taken from \citet{Asplund2009})\\
54 & log $\epsilon$(Eu) & $-$ & The logarithmic europium abundance of the star as given by the Reference\\
55 & EURDET & $-$ & Flag with ``D" if the europium abundance ratio ([Eu/Fe]) is detected by Reference and ``U" if\\
 &  &  & an upper limit by Reference and ``L" if a lower limit by Reference and ``X" if none is\\
 &  &  & detected by Reference\\
56 & Class & $-$ & The class for the star, as given by Table 1\\
57 & SNR & $-$ & The average Signal-to-Noise Ratio of the spectrum as given by the Reference\\
58 & Reference & $-$ & The Reference for the star\\
59 & $V_{\rm mag}$ Reference & $-$ & The Reference for the \textit{V} magnitude of the star\\
60 & Distance AGAMA & $-$ & The Reference for the distance used in AGAMA (StarHorse prioritized over BJ21 unless StarHorse\\
 &  &  & distance has relative error greater than 0.3, if both have a relative error greater than 0.3\\
 &  &  & we adopt no distance estimate)\\
61 & RV Reference & $-$ & The Reference for the RV\\
\enddata
\end{deluxetable*}

\startlongtable
\begin{deluxetable*}{c  l  l  l}
\tablecaption{Description of the Final Sample from the RPE Sample \label{tab:final_data_descript}}
\tablehead{\colhead{Column} & \colhead{Field} & \colhead{Unit} & \colhead{Description}}
\startdata
1 & Name & $-$ & The name of the star as given by the Reference\\
2 & Source ID & $-$ & The Gaia EDR3 Source ID of the star\\
3 & RA & (J2000) & The Right Ascension of the star given in hours:minutes:seconds\\
4 & DEC & (J2000) & The Declination of the star given in degrees:minutes:seconds\\
5 & $V_{\rm mag}$ & $-$ & The $V$ magnitude of the star as given by the $V_{\rm mag}$ Reference\\
6 & $G_{\rm mag}$ & $-$ & The Gaia $G$ mean magnitude of the star as given by the Gaia Source ID\\
7 & $G_{\rm BP} - G_{\rm RP}$ & $-$ & The Gaia BP $-$ RP color mean magnitude of the star as given by the Gaia Source ID\\
8 & $V_{\rm mag}$ (Gaia) & $-$ & The $V$ magnitude of the star as determined by the transformations from G mag to V mag using\\
 &  &  & $V$ $=$ $G$ $+$ $0.02704$ $-$ $0.01424*(\rm{BP}-\rm{RP})$ $+$ $0.2156*(\rm{BP}-\rm{RP})^2$ $-$\\
 &  &  & $0.01426(\rm{BP}-\rm{RP})^3$ given by \citet{Riello2021}\\
9 & RV & (km s$^{-1}$) & The radial velocity as given by RV Reference\\
10 & Error & (km s$^{-1}$) & The radial velocity error as given by RV Reference\\
11 & RV$_{\rm Gaia}$ & (km s$^{-1}$) & The radial velocity as given by the Gaia Source ID\\
12 & Error & (km s$^{-1}$) & The radial velocity error as given by the Gaia Source ID\\
13 & Parallax & (mas) & The parallax as given by the Gaia Source ID\\
14 & Error & (mas) & The parallax error as given by the Gaia Source ID\\
15 & Distance & (kpc) & The inverse parallax distance (1/Parallax)\\
16 & Error & (kpc) & The inverse parallax distance error (Parallax$_{error}$/(Parallax$^2$))\\
17 & Distance$_{\rm Corrected}$ & (kpc) & The corrected inverse parallax distance (1/(Parallax + 0.026)) based on \citet{Huang2021b}\\
18 & Error & (kpc) & The corrected inverse parallax distance error (Parallax$_{error}$/((Parallax + 0.026)$^2$))\\
 &  &  & based on \citet{Huang2021b}\\
19 & Relative Error & $-$ & The relative error of the corrected distance as given by Gaia\\
20 & Distance BJ21 & (kpc) & The 50 percentile distance as given by \citet{Bailer-Jones2021} based on the Gaia Source ID\\
21 & Error & (kpc) & The 50 percentile error as estimated by the 84 percentile distance and the 16 percentile\\
 &  &  & distance as given by \citet{Bailer-Jones2021} based on the Gaia Source ID ((dist84-dist16)/2)\\
22 & Relative Error & $-$ & The relative error of the 50 percentile distance as given by \citet{Bailer-Jones2021} based on\\
 &  &  & the Gaia Source ID\\
23 & Distance StarHorse & (kpc) & The 50 percentile distance as given by \citet{Anders2022} based on the Gaia Source ID\\
24 & Error & (kpc) & The 50 percentile error as estimated by the 84 percentile distance and the 16 percentile\\
 &  &  & distance as given by \citet{Anders2022} based on the Gaia Source ID ((dist84-dist16)/2)\\
25 & Relative Error & $-$ & The relative error of the 50 percentile distance as given by \citet{Anders2022} based on the\\
 &  &  & Gaia Source ID\\
26 & PM$_{\rm RA}$ & (mas yr$^{-1}$) & The proper motion in the Right Ascension as given by the Gaia Source ID\\
27 & Error & (mas yr$^{-1}$) & The proper motion error in the Right Ascension as given by the Gaia Source ID\\
28 & PM$_{\rm DEC}$ & (mas yr$^{-1}$) & The proper motion in the Declination as given by the Gaia Source ID\\
29 & Error & (mas yr$^{-1}$) & The proper motion error in the Declination as given by the Gaia Source ID\\
30 & Correlation Coefficient & $-$ & The correlation coefficient between the proper motion in Right Ascension and the proper motion\\
 &  &  & in Declination as given by the Gaia Source ID\\
31 & T$_{\rm eff}$ & (K) & The effective temperature of the star as given by the Reference\\
32 & log \textit{g} & (cgs) & The surface gravity of the star as given by the Reference\\
33 & [Fe/H] & $-$ & The metallicity of the star as given by (log $\epsilon$(Fe) - log $\epsilon$(Fe)$_\odot$)\\
 &  &  & (Solar value of 7.50 taken from \citet{Asplund2009})\\
34 & log $\epsilon$(Fe) & $-$ & The logarithmic iron abundance of the star as given by the Reference\\
35 & [C/Fe] & $-$ & The carbon abundance ratio of the star as given by (log $\epsilon$(C) - log\\
 &  &  & $\epsilon$(C)$_\odot$ - [Fe/H]) (Solar value of 8.43 taken from \citet{Asplund2009})\\
36 & log $\epsilon$(C) & $-$ & The logarithmic carbon abundance of the star as given by the Reference\\
37 & [C/Fe]$_{c}$ & $-$ & The carbon abundance ratio corrected for evolutionary effects from \citet{Placco2014b}\\
38 & AC$_{c}$ & $-$ & The absolute carbon corrected for evolutionary effects from \citet{Placco2014b} ([C/Fe]$_{c}$\\
 &  &  & + [Fe/H] + log $\epsilon$(C)$_\odot$) (Solar value of 8.43 taken from \citet{Asplund2009})\\
39 & CARDET & $-$ & Flag with ``D" if the carbon abundance ratio ([C/Fe]) is detected by Reference and ``U" if an\\
 &  &  & upper limit by Reference and ``L" if a lower limit by Reference and ``X" if none is detected\\
 &  &  & by Reference\\
40 & CEMP & $-$ & Flag as ``C" for Carbon-Enhanced Metal-Poor (CEMP) if [C/Fe]$_{c}$ $> +0.7$ and ``I" for\\
 &  &  & CEMP-intermediate if $+0.5 <$ [C/Fe]$_{c}$ $\leq +0.7$ and ``N" for Carbon-Normal if\\
 &  &  & [C/Fe]$_{c}$ $\leq +0.5$ and ``X" if there is no [C/Fe]$_{c}$ information\\
41 & [Mg/Fe] & $-$ & The magnesium abundance ratio of the star as given by (log $\epsilon$(Mg) - log\\
 &  &  & $\epsilon$(Mg)$_\odot$ - [Fe/H]) (Solar value of 7.60 taken from \citet{Asplund2009})\\
42 & log $\epsilon$(Mg) & $-$ & The logarithmic magnesium abundance of the star as given by the Reference\\
43 & MAGDET & $-$ & Flag with ``D" if the magnesium abundance ratio ([Mg/Fe]) is detected by Reference and ``U" if\\
 &  &  & an upper limit by Reference and ``L" if a lower limit by Reference and ``X" if none is\\
 &  &  & detected by Reference\\
44 & [Sr/Fe] & $-$ & The strontium abundance ratio of the star as given by (log $\epsilon$(Sr) - log\\
 &  &  & $\epsilon$(Sr)$_\odot$ - [Fe/H]) (Solar value of 2.87 taken from \citet{Asplund2009})\\
45 & log $\epsilon$(Sr) & $-$ & The logarithmic strontium abundance of the star as given by the Reference\\
46 & STRDET & $-$ & Flag with ``D" if the strontium abundance ratio ([Sr/Fe]) is detected by Reference and ``U" if\\
 &  &  & an upper limit by Reference and ``L" if a lower limit by Reference and ``X" if none is\\
 &  &  & detected by Reference\\
47 & [Y/Fe] & $-$ & The yttrium abundance ratio of the star as given by (log $\epsilon$(Y) - log\\
 &  &  & $\epsilon$(Y)$_\odot$ - [Fe/H]) (Solar value of 2.21 taken from \citet{Asplund2009})\\
48 & log $\epsilon$(Y) & $-$ & The logarithmic yttrium abundance of the star as given by the Reference\\
49 & YTTDET & $-$ & Flag with ``D" if the yttrium abundance ratio ([Y/Fe]) is detected by Reference and ``U" if an\\
 &  &  & upper limit by Reference and ``L" if a lower limit by Reference and ``X" if none is detected\\
 &  &  & by Reference\\
50 & [Ba/Fe] & $-$ & The barium abundance ratio of the star as given by (log $\epsilon$(Ba) - log\\
 &  &  & $\epsilon$(Ba)$_\odot$ - [Fe/H]) (Solar value of 2.18 taken from \citet{Asplund2009})\\
51 & log $\epsilon$(Ba) & $-$ & The logarithmic barium abundance of the star as given by the Reference\\
52 & BARDET & $-$ & Flag with ``D" if the barium abundance ratio ([Ba/Fe]) is detected by Reference and ``U" if an\\
 &  &  & upper limit by Reference and ``L" if a lower limit by Reference and ``X" if none is detected\\
 &  &  & by Reference\\
53 & [Eu/Fe] & $-$ & The europium abundance ratio of the star as given by (log $\epsilon$(Eu) - log\\
 &  &  & $\epsilon$(Eu)$_\odot$ - [Fe/H]) (Solar value of 0.52 taken from \citet{Asplund2009})\\
54 & log $\epsilon$(Eu) & $-$ & The logarithmic europium abundance of the star as given by the Reference\\
55 & EURDET & $-$ & Flag with ``D" if the europium abundance ratio ([Eu/Fe]) is detected by Reference and ``U" if\\
 &  &  & an upper limit by Reference and ``L" if a lower limit by Reference and ``X" if none is\\
 &  &  & detected by Reference\\
56 & Class & $-$ & The class for the star, as given by Table 1\\
57 & SNR & $-$ & The average Signal-to-Noise Ratio of the spectrum as given by the Reference\\
58 & Reference & $-$ & The Reference for the star\\
59 & $V_{\rm mag}$ Reference & $-$ & The Reference for the \textit{V} magnitude of the star\\
60 & Distance AGAMA & $-$ & The Reference for the distance used in AGAMA (StarHorse prioritized over BJ21 unless \\
 &  &  & StarHorse distance has relative error greater than 0.3, if both have a relative error\\
 &  &  & greater than 0.3 we adopt no distance estimate)\\
61 & RV Reference & $-$ & The Reference for the RV\\
62 & (v$_{\rm r}$,v$_{\phi}$,v$_{\rm z}$)  & (km s$^{-1}$) & The cylindrical velocities of the star as given by AGAMA\\
63 & Error & (km s$^{-1}$) & The cylindrical velocity errors of the star as given by Monte Carlo sampling through AGAMA\\
64 & (J$_{\rm r}$,J$_{\phi}$,J$_{\rm z}$) & (kpc km s$^{-1}$) & The cylindrical actions of the star as given by AGAMA\\
65 & Error & (kpc km s$^{-1}$) & The cylindrical action errors of the star as given by Monte Carlo sampling through AGAMA\\
66 & Energy & (km$^{2}$ s$^{-2}$) & The orbital energy of the star as given by AGAMA\\
67 & Error  & (km$^{2}$ s$^{-2}$) & The orbital energy error of the star as given by Monte Carlo sampling through AGAMA\\
68 & $r_{\rm peri}$ & (kpc) & The Galactic pericentric distance of the star as given by AGAMA\\
69 & Error & (kpc) & The Galactic pericentric distance error of the star as given by Monte Carlo sampling through\\
 &  &  & AGAMA\\
70 & $r_{\rm apo}$ & (kpc) & The Galactic apocentric distance of the star as given by AGAMA\\
71 & Error & (kpc) & The Galactic apocentric distance error of the star as given by Monte Carlo sampling through\\
 &  &  & AGAMA\\
72 & Z$_{\rm max}$ & (kpc) & The maximum height above, or below, the Galactic plane of the star as given by AGAMA\\
73 & Error & (kpc) & The maximum height above, or below, the Galactic plane error of the star as given by Monte\\
 &  &  & Carlo sampling through AGAMA\\
74 & Eccentricity & $-$ & The eccentricity of the star given by ($r_{\rm{apo}}$ $-$ $r_{\rm{peri}}$)/($r_{\rm{apo}}$ $+$\\
 &  &  & $r_{\rm{peri}}$) through AGAMA\\
75 & Error & $-$ & The eccentricity error of the star as given by Monte Carlo sampling through AGAMA\\
\enddata
\end{deluxetable*}

\startlongtable\begin{deluxetable*}{l  l  c  c  c  c  c  c  c  c  c  c  c}
\tabletypesize{\scriptsize}
\tablecaption{Identified CEMP stars and their Group Association \label{tab:cemp}}
\tablehead{\colhead{Name} & \colhead{Group} & \colhead{T$_{\rm eff}$ (K)} & \colhead{log $g$} & \colhead{[Fe/H]} & \colhead{[C/Fe]} & \colhead{[C/Fe]$_{c}$} & \colhead{AC$_{c}$} & \colhead{[Mg/Fe]} & \colhead{[Sr/Fe]} & \colhead{[Y/Fe]} & \colhead{[Ba/Fe]} & \colhead{[Eu/Fe]}}
\startdata
HE 0010$-$3422&I&$5400$&$3.100$&$-2.70$&+$1.84$&+$1.86$&$7.59$&+$0.26$&+$0.77$&+$0.93$&+$1.46$&+$1.64$\\[-0.05ex]
J004539.30$-$745729.4&I&$4947$&$2.010$&$-2.00$&+$0.93$&+$0.99$&$7.42$&$\dots$&+$0.83$&$\dots$&+$0.37$&+$0.55$\\[-0.05ex]
J005327.84$-$025316.8&I&$4370$&$0.560$&$-2.16$&+$0.40$&+$0.88$&$7.15$&+$0.48$&$-0.05$&$-0.31$&$-0.24$&+$0.39$\\[-0.05ex]
J005419.65$-$061155.4&I&$4707$&$1.030$&$-2.32$&+$0.50$&+$0.98$&$7.09$&+$0.52$&+$0.26$&$-0.25$&$-0.21$&+$0.59$\\[-0.05ex]
J010839.60$-$285701.0*&II&$4843$&$1.170$&$-2.87$&+$0.10$&+$0.76$&$6.32$&+$0.41$&+$0.41$&$\dots$&$-0.52$&+$0.45$\\[-0.05ex]
J014908.00$-$491143.0*&II&$4556$&$0.720$&$-3.08$&+$0.07$&+$0.81$&$6.16$&+$0.63$&$-0.01$&$\dots$&$-0.39$&+$0.35$\\[-0.05ex]
J015656.33$-$140210.9&I&$4622$&$1.090$&$-2.08$&+$0.37$&+$0.83$&$7.18$&+$0.37$&+$0.10$&$-0.23$&$-0.11$&+$0.76$\\[-0.05ex]
J022124.00$-$405648.0&II&$4797$&$0.940$&$-2.52$&+$0.20$&+$0.83$&$6.74$&$\dots$&$-0.18$&$\dots$&$-0.18$&+$0.46$\\
\enddata
\tablecomments{The CEMP Groups are given in \citet{Yoon2016}.}
\tablecomments{Stars with a '*' at the end of their name denote that they did not make the RPE Final Sample.}
\end{deluxetable*}

\startlongtable
\begin{deluxetable*}{l  c  c  c  c  c  c  c  c}
\tablecaption{CDTGs Identified by \HDBSCAN~ for the $r$-I Sample \label{tab:cluster_results_rI_stub}}
\tablehead{\colhead{NAME} & \colhead{[Fe/H]} & \colhead{[C/Fe]$_{c}$} & \colhead{[Mg/Fe]} & \colhead{[Sr/Fe]} & \colhead{[Y/Fe]} & \colhead{[Ba/Fe]} & \colhead{[Eu/Fe]}}
\startdata
\multicolumn{8}{c}{CDTG-$1$-$r$I} \\
\multicolumn{8}{c}{Structure: Metal-Weak Thick Disk (MWTD)} \\
\multicolumn{8}{c}{Group Assoc: C-3: \citet{Malhan2022}} \\
\multicolumn{8}{c}{Stellar Assoc: 2MASS J00452379-2112161 (CDTG-16: \citealt{Gudin2021})} \\
\multicolumn{8}{c}{Stellar Assoc: J223454.50-660517.0 (DTG-132: \citealt{Shank2022b}) } \\
\multicolumn{8}{c}{Globular Assoc: No Globular Associations} \\
\multicolumn{8}{c}{Dwarf Galaxy Assoc: No Dwarf Galaxy Associations} \\
161116002201265 & $-0.90$ & $+0.50$ & $\dots$ & $\dots$ & $+0.10$ & $-0.09$ & $+0.54$ \\
\enddata
\tablecomments{$\boldsymbol{\mu}$ and $\boldsymbol{\sigma}$ represent the biweight estimates of the location and scale for the abundances in the CDTG.}\tablecomments{This table is a stub; the full table is available in the electronic edition.}\end{deluxetable*}

\startlongtable
\begin{deluxetable*}{l  r  c  c  r  r}
\tablecaption{CDTG Dynamical Parameters Determined by \AGAMA~ for the $r$-I Sample \label{tab:cluster_orbit_rI}}
\tablehead{\colhead{Cluster} & \colhead{$N$ Stars} & \colhead{($\langle$v$_{\text{r}}\rangle$,$\langle$v$_{\phi}\rangle$,$\langle$v$_{\text{z}}\rangle$)} & \colhead{($\langle$J$_{\text{r}}\rangle$,$\langle$J$_{\phi}\rangle$,$\langle$J$_{\text{z}}\rangle$)} & \colhead{$\langle$E$\rangle$} & \colhead{$\langle$ecc$\rangle$}\\
\colhead{} & \colhead{} & \colhead{($\sigma_{\langle\text{v}_{\text{r}}\rangle}$,$\sigma_{\langle\text{v}_{\phi}\rangle}$,$\sigma_{\langle\text{v}_{\text{z}}\rangle}$)} & \colhead{($\sigma_{\langle\text{J}_{\text{r}}\rangle}$,$\sigma_{\langle\text{J}_{\phi}\rangle}$,$\sigma_{\langle\text{J}_{\text{z}}\rangle}$)} & \colhead{$\sigma_{\langle\text{E}\rangle}$} & \colhead{$\sigma_{\langle\text{ecc}\rangle}$}\\
\colhead{} & \colhead{} & \colhead{(km s$^{-1}$)} & \colhead{(kpc km s$^{-1}$)} & \colhead{(10$^{5}$ km$^{2}$ s$^{-2}$)} & \colhead{}}
\startdata
CDTG-$1$-$r$I & $18$ & ($-17.4$,$91.9$,$-25.5$) & ($239.7$,$648.7$,$162.3$) & $-1.768$ & $0.573$ \\
 & & ($74.9$,$25.4$,$59.9$) & ($52.3$,$216.6$,$13.1$) & $0.054$ & $0.096$ \\
CDTG-$2$-$r$I & $18$ & ($33.9$,$21.8$,$-3.6$) & ($580.1$,$157.5$,$55.9$) & $-1.808$ & $0.898$ \\
 & & ($104.1$,$12.1$,$51.1$) & ($59.6$,$83.1$,$25.3$) & $0.027$ & $0.047$ \\
CDTG-$3$-$r$I & $17$ & ($1.4$,$181.2$,$21.4$) & ($89.5$,$1446.0$,$136.0$) & $-1.596$ & $0.267$ \\
 & & ($65.9$,$25.8$,$69.6$) & ($57.7$,$83.9$,$31.9$) & $0.032$ & $0.093$ \\
CDTG-$4$-$r$I & $15$ & ($-27.0$,$72.3$,$5.2$) & ($164.6$,$295.6$,$243.3$) & $-1.936$ & $0.625$ \\
 & & ($59.9$,$50.3$,$101.3$) & ($75.4$,$199.8$,$32.5$) & $0.027$ & $0.201$ \\
\enddata
\end{deluxetable*}

\include{Tables/cluster_mean_rI_table}

\startlongtable
\begin{deluxetable*}{l  c  c  c  c  c  c  c  c}
\tablecaption{CDTGs Identified by \HDBSCAN~ for the $r$-II Sample \label{tab:cluster_results_rII_stub}}
\tablehead{\colhead{NAME} & \colhead{[Fe/H]} & \colhead{[C/Fe]$_{c}$} & \colhead{[Mg/Fe]} & \colhead{[Sr/Fe]} & \colhead{[Y/Fe]} & \colhead{[Ba/Fe]} & \colhead{[Eu/Fe]}}
\startdata
\multicolumn{8}{c}{CDTG-$1$-$r$II} \\
\multicolumn{8}{c}{Structure: Metal-Weak Thick Disk (MWTD)} \\
\multicolumn{8}{c}{Group Assoc: CDTG-3: \citet{Gudin2021}} \\
\multicolumn{8}{c}{Group Assoc: CDTG-25: \citet{Gudin2021}} \\
\multicolumn{8}{c}{Group Assoc: DTG-3: \citet{Shank2022b}} \\
\multicolumn{8}{c}{Group Assoc: DTG-36: \citet{Shank2022b}} \\
\multicolumn{8}{c}{Group Assoc: DTG-62: \citet{Shank2022b}} \\
\multicolumn{8}{c}{Group Assoc: DTG-82: \citet{Shank2022b}} \\
\enddata
\tablecomments{$\boldsymbol{\mu}$ and $\boldsymbol{\sigma}$ represent the biweight estimates of the location and scale for the abundances in the CDTG.}\tablecomments{This table is a stub; the full table is available in the electronic edition.}\end{deluxetable*}

\startlongtable
\begin{deluxetable*}{l  r  c  c  r  r}
\tablecaption{CDTG Dynamical Parameters Determined by \AGAMA~ for the $r$-II Sample \label{tab:cluster_orbit_rII}}
\tablehead{\colhead{Cluster} & \colhead{$N$ Stars} & \colhead{($\langle$v$_{\text{r}}\rangle$,$\langle$v$_{\phi}\rangle$,$\langle$v$_{\text{z}}\rangle$)} & \colhead{($\langle$J$_{\text{r}}\rangle$,$\langle$J$_{\phi}\rangle$,$\langle$J$_{\text{z}}\rangle$)} & \colhead{$\langle$E$\rangle$} & \colhead{$\langle$ecc$\rangle$}\\
\colhead{} & \colhead{} & \colhead{($\sigma_{\langle\text{v}_{\text{r}}\rangle}$,$\sigma_{\langle\text{v}_{\phi}\rangle}$,$\sigma_{\langle\text{v}_{\text{z}}\rangle}$)} & \colhead{($\sigma_{\langle\text{J}_{\text{r}}\rangle}$,$\sigma_{\langle\text{J}_{\phi}\rangle}$,$\sigma_{\langle\text{J}_{\text{z}}\rangle}$)} & \colhead{$\sigma_{\langle\text{E}\rangle}$} & \colhead{$\sigma_{\langle\text{ecc}\rangle}$}\\
\colhead{} & \colhead{} & \colhead{(km s$^{-1}$)} & \colhead{(kpc km s$^{-1}$)} & \colhead{(10$^{5}$ km$^{2}$ s$^{-2}$)} & \colhead{}}
\startdata
CDTG-$1$-$r$II & $23$ & ($-28.6$,$163.0$,$-6.9$) & ($117.8$,$1241.4$,$51.4$) & $-1.671$ & $0.338$ \\
 & & ($67.2$,$25.6$,$47.9$) & ($94.5$,$150.1$,$34.4$) & $0.030$ & $0.120$ \\
CDTG-$2$-$r$II & $18$ & ($-3.6$,$17.6$,$-17.9$) & ($929.9$,$146.9$,$59.3$) & $-1.612$ & $0.929$ \\
 & & ($198.8$,$11.0$,$37.3$) & ($104.4$,$108.8$,$31.9$) & $0.050$ & $0.038$ \\
CDTG-$3$-$r$II & $15$ & ($-73.5$,$15.4$,$29.7$) & ($694.7$,$128.6$,$158.2$) & $-1.682$ & $0.923$ \\
 & & ($117.3$,$26.2$,$97.6$) & ($53.5$,$202.8$,$42.1$) & $0.023$ & $0.059$ \\
CDTG-$4$-$r$II & $12$ & ($-312.0$,$19.7$,$0.5$) & ($1460.5$,$159.9$,$78.6$) & $-1.383$ & $0.933$ \\
 & & ($166.5$,$46.2$,$80.3$) & ($159.9$,$343.4$,$30.3$) & $0.034$ & $0.044$ \\
\enddata
\end{deluxetable*}

\include{Tables/cluster_mean_rII_table}

\bibliography{main}{}
\bibliographystyle{aasjournal}

\end{document}